\documentclass[a4paper]{cedram-smai-jcm}

\usepackage[utf8]{inputenc}
\usepackage{mathtools}
\usepackage{gensymb}
\usepackage{subfig}
\usepackage{tikz}
\usepackage{pgfplots}
\pgfplotsset{compat=1.14}
\usetikzlibrary{shapes,positioning,arrows.meta}
\usepackage{jlcode}
\usepackage{cleveref}

\DeclareMathOperator{\spn}{span}
\DeclareMathOperator{\divergence}{div}
\DeclareMathOperator{\ind}{ind}
\DeclareMathOperator{\sign}{sign}
\DeclareMathOperator{\Ret}{Ret}
\DeclareMathOperator{\trace}{trace}
\DeclareMathOperator{\rem}{rem}
\DeclareMathOperator{\round}{round}

\newcommand{\R}{\mathbb{R}}
\newcommand{\lf}{\ell}

\newcommand{\dd}{\mathrm{d}}
\newcommand{\density}{c}
\newcommand{\velocity}{\mathbf{v}}
\newcommand{\flow}{\mathbf{F}}
\newcommand{\diffusivity}{\nu}
\newcommand{\difftensor}{\mathbf{D}}
\newcommand{\normal}{\mathbf{n}}

\title[Fast computation of coherent Lagrangian vortices]{Fast and robust computation of coherent Lagrangian vortices\\
on very large two-dimensional domains}

\author[D. Karrasch]{\firstname{Daniel} \lastname{Karrasch}}
\address{Department of Mathematics, Technische Universität München, Garching bei München, Germany}
\email{karrasch@ma.tum.de}
\thanks{This work is supported by the Priority Programme SPP 1881 Turbulent
Superstructures of the Deutsche Forschungsgemeinschaft.}

\author[N. Schilling]{\firstname{Nathanael} \lastname{Schilling}}
\address{Department of Mathematics, Technische Universität München, Garching bei München, Germany}

  

\begin{document}

\begin{abstract}
We describe a new method for computing coherent Lagrangian vortices
in two-dimensional flows according to any of the following approaches:
\emph{black-hole vortices} \cite{Haller2013a}, \emph{objective Eulerian Coherent Structures (OECSs)}
\cite{Serra2016}, \emph{material barriers to diffusive transport} \cite{Haller2018,Haller2019}, and
\emph{constrained diffusion barriers} \cite{Haller2019}.
The method builds on ideas developed previously in \cite{Karrasch2015}, but our
implementation alleviates a number of shortcomings and allows for the fully
automated detection of such vortices on unprecedentedly challenging real-world
flow problems, for which specific human interference is absolutely infeasible.
Challenges include very large domains and/or parameter spaces.
We demonstrate the efficacy of our method in dealing with such challenges
on two test cases: first, a parameter study of a turbulent flow, and second,
computing material barriers to diffusive transport in the global ocean.
\end{abstract}

\maketitle


\section{Introduction}

\emph{Coherent Lagrangian vortices (CLVs)} play an important role in the transport and mixing of
passive, possibly diffusive, scalar quantities in fluid flows. Such structures can be found in flows
living on a wide variety of scales \cite{Huhn2015a,Hadjighasem2017,Abernathey2018}.

Over the past decade, a variety of modeling approaches have been developed to characterize CLVs.
Intuitively, CLVs are viewed as material structures that sustain a non-filamenting boundary under advection by the flow
\cite{Haller2012,Haller2013a,Froyland2015a}; or material structures that resist
leakage of a diffusive passive scalar in an advection--diffusion process
\cite{Karrasch2016b,Haller2018,Haller2019}. Data-driven approaches view CLVs as collections of trajectories that stay together under the
motion of the flow \cite{Allshouse2012,Froyland2015,Hadjighasem2016,Banisch2017,Padberg-Gehle2017}.
Related Eulerian approaches view coherent sets as space-time structures that do
not mix much with their spatial neighborhood \cite{Froyland2010,Froyland2013}.
While these approaches intuitively target the same observed phenomenon, they
often yield different structures when applied to the same flow \cite{Hadjighasem2017}.

To date, only few methods have been successfully applied to realistic flow problems. See, for instance,
\cite{Froyland2010a,Haller2013a,Karrasch2015,Hadjighasem2016b,Hadjighasem2017,Serra2017b,Froyland2019}
for studies of ``medium'' complexity (in that either the domain is not too large or the number of
known/expected structures is low), and \cite{Abernathey2018} for the only---to the best of our knowledge---large-scale study.
The reasons for this lack of realistic applications are manifold: (i) most methods are not fully automated or even automizable,
cf.~\cite{Hadjighasem2017}, (ii) some methods intrinsically do not scale well with the
size of the domain, the number of expected coherent structures, or the number of
tracked trajectories; and/or (iii) there are no performant and robust implementations available.

Our aim in this paper is to report on our progress towards bridging the gap
between large-scale applications and those methods that subsume
the ``geodesic vortices'' class: \emph{black-hole vortices} \cite{Haller2013a}, \emph{objective Eulerian Coherent Structures (OECSs)}
\cite{Serra2016}, and \emph{material barriers to diffusive transport} \cite{Haller2018,Haller2019}. The algorithms are
developed as part of the open-source \texttt{CoherentStructures.jl} project. These
methods in principle scale well with the size of the domain, but existing
implementations related to the publications \cite{Onu2015,Karrasch2015,Hadjighasem2016b,Serra2017}
failed to fully leverage this; we therefore had to make significant
conceptual and implementation modifications.

Conceptually, our work is based on the index-theory-based methodology developed
in \cite{Karrasch2015}, whose implementation was a mixture of methods
implemented in \cite{Farazmand2014a,Onu2015,Hadjighasem2016b}. In
\cite{Serra2017}, Serra \& Haller identified (i) the detection of tensor field
singularities (points of repeated tensor eigenvalues) of tensor fields and (ii) the
identification of their topological type as major computational bottlenecks in the
implementation of \cite{Karrasch2015}. Moreover, these steps required a number of
parameters whose choice had---at times---unpredictable impact on the
computational outcome. As an alternative, they derived an automated method for
computing geodesic vortices based on the geometry of the underlying geodesic
flow. On the upside, their approach (i) does not require singularity detection and
type identification at all, and (ii) is designed to, in principle, not miss any coherent vortices at a
given computational accuracy and spatial resolution. On the downside,
however, (i) the gradient of the underlying computed tensor field is required, (ii) the
computation is performed on the whole domain at once without a localization option;
and (iii) the currently available implementation is not performant.

In our implementation, we have carefully addressed the issues raised by
\cite{Serra2017} regarding the implementation of \cite{Karrasch2015}. Specifically,
we have improved a number of aspects related to the (inherently robust) topological
index-theory-based methods. In the spirit of discrete differential geometry, we now discretize the
tensor index computation in a manner such that important properties are preserved, and these
properties are exploited efficiently.
The main reason why we argue it is worth improving on the index-based approach is
that it allows for the identification of a comparatively small number of candidate regions that each potentially contain
a CLV, around which one may then restrict subsequent computations. This in particular allows for
straightforward parallelization. Ultimately, flow problems of high complexity become manageable.

Implemented in the modern and performant programming language \texttt{Julia}
\cite{Bezanson2017}, our package is able to find geodesic vortices on domains of
unprecedented size and orders of magnitude faster than what is the current state of
the art. Specifically, we demonstrate our code (i) on a parameter-dependent
turbulent flow and (ii) in a global ocean surface
simulation using a computational grid of tens of millions of points. The required
computational power does not exceed what is available on an ordinary work station
or a modern desktop machine. While there remains room for further improvement,
this shows that it is possible to effectively compute CLVs in very large-scale 2D
flows and/or to perform extensive parameter studies on medium-sized domains.

This paper is organized as follows. In \cref{sec:advectiondiffusion}, we recall the
mathematical framework of \cite{Haller2018} that introduced the concept of
``material barriers (to diffusive transport)'' as an instance of the methods falling into
the category of geodesic vortices. These are then summarized together with a
generic computational approach in \cref{sec:geodesic_vortices}.
\Cref{compapproach} is devoted to the description of our computational approach
based on index theory for planar line fields. For convenience, we have collected
related facts in \cref{app:index_theory}. Finally, we demonstrate the outstanding
capabilities of our implementation on two non-trivial applications in
\cref{sec:applications}: a parameter study based on a minimal two-dimensional
turbulence simulation on the torus, and a global ocean surface velocity simulation.

\section{Background}

\subsection{Mathematical setting}
\label{sec:advectiondiffusion}
We now recall the theory related to material barriers to diffusive transport\cite{Haller2018},
as the most recent instance of a method that fits into the geodesic vortex framework. This is also the
method we use in the examples in \cref{sec:applications}.
Here the setting is a time-dependent
incompressible fluid velocity field $\velocity\colon U \times \mathcal{T} \to \R^2$,
where $U$ is an open, simply connected subset of $\R^2$ and $\mathcal{T}$ is a finite time interval.
A passive scalar $u$, i.e., a scalar quantity that does not affect the velocity field,
undergoes \emph{advection-diffusion} if it satisfies the partial differential equation (PDE)
\begin{equation}\label{eq:ade}
\partial_t \density+ \divergence (\density \cdot \velocity) = \diffusivity\divergence\nabla \density = \diffusivity\Delta \density\,.
\end{equation}
In words, the density $\density$ is carried by the fluid and diffuses isotropically. The
inclusion of anisotropic and/or spatially inhomogeneous and time-dependent
diffusion is straightforward, but omitted here for ease of presentation.
\Cref{eq:ade} models the evolution of a range of physically relevant quantities,
including concentrations of dissolved substances (like salinity and temperature) and
vorticity in the 2D Navier--Stokes equations. Strictly speaking, none of these three examples is
passive, but certainly temperature and salinity can be regarded as such over time
scales of a few weeks or even months.
Furthermore, \cref{eq:ade} on $\mathbb R^2$ can be interpreted
as the \emph{Fokker--Planck/Kolmogorov forward equation} of the stochastic differential equation
\[
dX_t = \velocity(X_t,t)dt + \sqrt{2\diffusivity}dW_t\,,
 \]
provided that $\velocity$ satisfies certain regularity assumptions; cf.~\cite{Karatzas1991}.

The initial value $\density(\cdot,0) = \density_0$ uniquely defines the solution of \cref{eq:ade} given
appropriate boundary conditions on $\partial U \times \mathcal{T}$.
The value $\diffusivity > 0$ is the \emph{diffusivity} (or the inverse Péclet number in
the non-dimensionalized form) and is very small in many applications. In the
absence of diffusion ($\epsilon = 0$), $\density$ is conserved and transported
along the characteristics of the velocity field $\velocity$. Characteristics $x_t$ satisfy the
ordinary differential equation (ODE) $\frac{\dd}{\dd t}x_t = \velocity(x_t,t)$.
We denote by $\flow_{t_0}^t$ the flow-map for this ordinary differential equation,
i.e., $\flow_{t_0}^t(p)$ corresponds to the time-$t$ solution of the ODE with initial value $x_{t_0} = p$.

By definition, \emph{Lagrangian} (or \emph{material} structures) are
\emph{invariant} under the flow $t \mapsto \flow_{t_0}^t$. Hence, the flow map can
be used to \emph{define} Lagrangian coordinates in space by labelling a
spatialtemporal point $(x,t)$ using the fluid ``particle'' $p$ that occupies $x$ at time $t$.
Clearly, in Lagrangian coordinates there is no advective transport, moreover with this change of coordinates
the advection--diffusion equation \eqref{eq:ade} takes the form of a pure diffusion equation \cite{Press1981,Thiffeault2003,Karrasch2016b,Haller2018}
\begin{equation}\label{eq:lade}
\partial_t \tilde\density =  \diffusivity \divergence\left(\mathbf{D}_{t_0}^t \nabla\tilde\density\right),
\end{equation}
where $\mathbf{D}_{t_0}^t(p) = (D\flow_{t_0}^t(p,t))^{-1} (D\flow_{t_0}^t(p,t))^{-\top}$ and $\tilde\density(p,t) = \density(\flow_{t_0}^t(p),t)$
are, respectively, the diffusion tensor and the scalar density in Lagrangian coordinates. For notational simplicity, we omit the tilde in the notation of the scalar
density function in Lagrangian coordinates.
This coordinate change allows for the separation of the reversible effects of advection from the
irreversible effects of the combined advection and diffusion. Note that the Lagrangian diffusion tensor field $\difftensor$ is both $t$- and $p$-dependent.

In this framework, material barriers to diffusive and stochastic transport have been
defined in \cite{Haller2018} as material surfaces which extremize diffusive transport
over the finite observation time interval. There, it is shown that in two-dimensional
flows, the diffusive transport through a one-dimensional material manifold
$\Gamma$ is given in leading order (with respect to diffusivity $\diffusivity$) by
\begin{align}
\int_\Gamma \left\langle\nabla \density_0, \mathbf{T}_{t_0}^t\normal\right\rangle\,\mathrm{d}A\,,
\end{align}
where $\mathbf{T}_{t_0}^t$ is the \emph{transport tensor field}, defined as the
time-average of the Lagrangian diffusion tensor fields,
i.e.~$\mathbf{T}_{t_0}^t = \frac{1}{t-t_0}\int_{t_0}^t \mathbf{D}_{t_0}^\tau\,\mathrm{d}\tau$,
$\normal$ is the outward-pointing normal, and $\mathrm{d}A$ is the canonical (Euclidean) surface measure.
After normalizing by the length
\footnote{The material barrier theory applies to higher dimensions, but the implementation for 2 dimensions does not generalize easily to 3 or more spatial dimensions.}
of $\Gamma$ and choosing a most ``diffusion-prone'' distribution of $\density_0$, a functional
on closed curves is found whose stationary points are null-geodesics
of an indefinite metric tensor field. This fits nicely into the ``geodesic vortex''
framework described in the next section. As a by-product, the trace of the transport
tensor, $\trace(\mathbf{T})$, coined \emph{diffusion barrier strength (DBS)}, is a
diagnostic field whose logarithm we will use for visualization purposes as a scalar
background field in this work.

\subsection{Coherent Lagrangian vortices as null-geodesics}
\label{sec:geodesic_vortices}

Mathematically speaking, the ``geodesic vortex'' approach consists of the
computation of closed null-geodesics of a (possibly
indefinite) metric tensor field that has undergone a parameter-dependent shift.
Three different vortex approaches can be formulated
in this setting:
\begin{enumerate}
\item the ``black hole vortex'' approach \cite{Haller2013a}, which seeks stationary curves of
a functional related to stretching;
\item the ``objective Eulerian coherent structures'' (OECS) approach \cite{Serra2016}, which seeks stationary
curves of a functional related to instantaneous stretching (i.e., strain);
\item the ``material barriers to diffusive transport'' \cite{Haller2018,Haller2019} approach described in the previous section.
\end{enumerate}
In these three cases, the tensor field comes from, respectively, (a) the Cauchy--Green strain tensor
$\mathbf{C}_{t_0}^t = (D\flow_{t_0}^t)^\top D\flow_{t_0}^t$, (b) the rate of strain tensor $\mathbf{S}_{t_0}$ (i.e., the
symmetric part of the velocity gradient $D\velocity(\cdot,t_0)$), and (c) the transport tensor $\mathbf{T}_{t_0}^t$ defined in the previous section.
In the following, we will use the generic $\mathbf{T}$ to denote any of these tensor fields.

The geodesic vortex approach seeks to find closed null-geodesic curves of
$\mathbf{T}-\lambda\mathbf{I}$; here the (real) parameter $\lambda$ is taken from a physically motivated range.
Recall that null-geodesics are smooth curves $\gamma$ that have ``zero length''
when measured in the indefinite metric $\mathbf{T} - \lambda \mathbf{I}$,
i.e. $\gamma' \cdot (\mathbf{T}-\lambda\mathbf{I})\gamma' = 0$. It is readily verified
that they can be computed as integral curves of
\begin{equation}\label{eq:eta}
\eta_\lambda^{\pm} = \sqrt{\frac{\lambda_2 - \lambda}{\lambda_2 - \lambda_1}}\xi_1 \pm \sqrt{\frac{\lambda-\lambda_1}{\lambda_2-\lambda_1}}\xi_2\,,
\end{equation}
where $\lambda_1 \leq \lambda_2$ are eigenvalues of $\mathbf{T}$ and $\xi_1,\xi_2$ are
corresponding normalized eigenvectors. This is derived from the fact \cite{Haller2013a} that
null-geodesics $\gamma$ have uniform $\mathbf{T}$-strain along themselves, i.e.,
along $\gamma$ one has
\begin{equation}\label{eq:lamcons}
\sqrt{\frac{\gamma'\cdot \mathbf{T} \gamma'}{\gamma' \cdot \gamma'}} = \lambda\,.
\end{equation}

As such, null-geodesics are closed integral curves of a planar line field, to which a corresponding
index theory applies; cf.~\cref{app:index_theory,app:applications}. Analogously to index theory for
planar vector fields, closed null-geodesics have index 1 relative to their inducing line field.
As defined formally in \cref{app:index_theory}, we will (i) assign an index (relative to
the line-field) to regions that are sets whose boundary forms a closed Jordan curve, or that are finite disjoint unions of such sets and (ii) refer
to all such regions with index 1 as \emph{elliptic regions}.

While the elliptic regions we initially identify will not have geodesic vortices as their boundaries per se,
we aim to find geodesic vortices that are homotopic to the elliptic regions found. The identification
of elliptic regions is the first step we use for the computation of geodesic vortices (this builds on \cite{Haller2013a,Karrasch2015}).

\section{The computational approach}\label{compapproach}

In this section, we give details of our implementation approach.
The structure of this section closely resembles the high-level structure
of our implementation provided in \texttt{CoherentStructures.jl}.
Our computational approach, at the highest level, consists of three steps:
\begin{enumerate}
\item Identify certain elliptic regions as candidate regions near geodesic vortices, based on index theoretical methods applied to the first eigenvector of $T$;
cf.~\cref{app:index_theory,app:applications}.
for the foundational theoretical aspects.

\item For each identified elliptic region, localize tensor field data to a neighborhood of the region.
\item Compute closed orbits (i.e., geodesic vortices) by a shooting method in this neighborhood.
\end{enumerate}

We describe each step in more detail in the following sections.

\subsection{Identification of elliptic regions}\label{sec:elliptic}

In a computational setting, we know the values of the tensor field $\mathbf{T}$ and,
hence, its subdominant eigenvector field $\lf = \xi_1$, only at a finite number of
points, say, the nodes of a polygonal triangulation/mesh $\mathcal{P}$. A candidate
elliptic region $R$ will be the union of a finite set of polygonal faces $P_1,\dots,P_k$
from $\mathcal{P}$. Such regions are identified in three steps:
\begin{enumerate}
\item Compute indices of every mesh face in a fast and robust manner solely from
the given tensor data, without interpolation; cf.~\cref{ssec:polygon}.
\item Suitably merge mesh faces into regions with stable index, and extract those that are elliptic; cf.~\cref{ssec:combination}.
\item Optionally, do further merging to obtain larger elliptic regions.
\end{enumerate}
The second and third steps are necessary because generically, the only structurally stable
singularities occurring in regular tensor fields have index $\pm\frac12$ \cite{Delmarcelle1994}.
Therefore, unless treating a degenerate or artificial tensor field, sufficiently small single polygonal
regions are not elliptic. To identify elliptic regions, we merge nearby polygonal
faces with non-vanishing indices in step (ii). Consistently with the additivity of the
index under curve composition/region merging (see \cref{app:index_theory}), we
add indices of polygonal cells when they are merged.

So far we have not yet specified criteria for deciding which polygonal regions should be merged.
We argue that a reasonable, robust criterion is to require that a candidate region's
index shall not change when (further) enlarged by a specified radius $r>0$.
This is captured by the following definition.

\begin{definition}
We say that a region $R$ is \emph{$r$-stable} (relative to $\lf$), if the set $R_s \coloneqq \lbrace x \in \Omega;~d(x,R) \leq  s\rbrace$
has the same index (relative to $\lf$) as $R$ for any $0 \leq s \leq r$.
\end{definition}

Clearly, if a region $R$ is $r$-stable, then within the $r$-vicinity of its boundary all polygonal faces have index 0.
Candidate regions for being homotopic to geodesic vortices are taken to be those that are minimal (with respect to inclusion)
unions of polygons that are $r$-stable and elliptic.
We note that in some cases the assumption of minimality is too strong;
as observed in \cite{Karrasch2015}, it is common for large geodesic vortices to bound
exactly two $r$-stable regions, each of wedge-type (i.e., index $\frac{1}{2}$).
In order to include also these elliptic regions, we additionally introduce
a number of ways to merge multiple $r$-stable regions that have indices summing to $1$.

Before giving a description of the details of steps (i)--(iii),
we summarize that our identification method
is fast, works directly on the line field data at mesh points, and can be used on
unstructured, irregular meshes/grids.
Moreover, it is unnecessary to choose pointwise
orientations for the line field, or to use ad-hoc heuristics for singularity type
classification. Robustness against local computational errors is achieved by
automatically choosing contours large enough so that any enlargement of the
contour (up to a specified size given by $r$) yields the same result. This is the only
parameter required by the (indirect) singularity detection method.

\subsubsection{Step (i): Calculating indices}
\label{ssec:polygon}

Assume we have a polygonal mesh $\mathcal{P}$ on $\Omega$ consisting of
vertices $\mathcal{V}$, edges $\mathcal{E}$, and polygonal faces $\mathcal{F}$. The vertices are points at which
the value of the line field $\lf$ is known. Since we are working in a discrete setting,
the natural curves to consider for the computation of indices are concatenations of
edges in $E$. To this end, let $\gamma$ be a simple closed Jordan curve along $n$ edges of the mesh, i.e., passing through the vertices $v_1,\ldots, v_n,v_{n+1} = v_1$ along the edges $e_i=(v_i,v_{i+1})$  and enclosing a union of polygons;
cf.~\cref{fig:combining}.

Since we know the value of the line field $\lf$ only at the vertices, we need to
approximate the curve $\lf \circ \gamma\colon [1,n+1] \rightarrow \mathbb P^1$ based on those
values in order to approximate $\theta$ used in \cref{def:lf_index}; \cref{app:index_theory}.
We cannot apply \cref{def:lf_index} to the discrete case directly as the angles
$\theta_i$ between $\lf(v_i)$ and the $x$-axis are determined only up to a
multiple of $\pi$. We follow \cite{Tricoche2004} and choose
$\theta_i$ such that the angle difference $\Delta_i\coloneqq\theta_{i+1}-\theta_i$
between subsequent angle representations is minimal modulo $\pi$ for $i=1,\dots,n$.
This is achieved by setting
\begin{equation}\label{eq:angleupdate}
\Delta_i \coloneqq \rem(\alpha_{i+1}-\alpha_i, \pi) = (\alpha_{i+1}-\alpha_i) - \pi\round\left(\frac{\alpha_{i+1}-\alpha_i}{\pi}\right)\,,
\end{equation}
where $\alpha_i$ is \emph{any} angle representation of $\lf(v_i)$. The index is then approximated by
\[
\ind_\lf(\gamma) \coloneqq \frac{1}{2\pi}\left( \theta_n - \theta_1 \right) = \frac{1}{2\pi} \sum_{i=1}^{n} \left(\theta_{i+1} - \theta_i\right) = \frac{1}{2\pi} \sum_{i=1}^{n} \Delta_i\,,
\]
where the right hand side can be viewed as a discretization of the integral representation of
the index in \cref{eq:lfindex}. We will refer to $\ind_\lf$ as the ``computed index'' whenever
we wish to explicitly distinguish this from the true index, though we will not always make the distinction.
We never have to pick an orientation
$\theta_i$ for the line field at the vertices, but only compute the angle updates
$\Delta_i$ via \cref{eq:angleupdate} from any angle representation
$\alpha_i$; the latter is usually obtained by calling the \texttt{arctan} function on the
line field components. Moreover, the value $\Delta_i$ only depends on the (directed)
edge $e_i$ and not on the rest of $\gamma$. Hence, $\Delta$ can be established
as a function on the set of edges $\mathcal{V}$. This method is used in \cite{Tricoche2004} for line field simplification by merging of
singularities, where it is shown that for linear line fields this
approach yields the correct index of an interpolated line-field on triangular meshes -- even though
these resolve the angle function $\theta$ by as few as three values. If $\gamma$ encloses
a region $R$ (and is positively oriented), define $\ind_\lf(R)  = \ind_\lf(\gamma)$.

\begin{figure}
\centering
\subfloat[A triangle in a mesh, with vertices $v_1$, $v_2$, $v_3$ and edges $e_1$, $e_2$, $e_3$ labelled.]{\begin{tikzpicture}[]
\begin{axis}[width = {0.5\textwidth}, height={0.41\textwidth}, axis lines=none,
ylabel = {},
xmin = {-0.1}, xmax = {3.1}, ymin = {-0.1}, ymax = {3.2}, 
xlabel = {}, unbounded coords=jump,
xshift = 0.0mm, yshift = 0.0mm,
]

\addplot+ [color = black, line width = 1, solid, mark = none,
fill = {rgb,1:red,0.00000000;green,0.39215686;blue,0.00000000},
forget plot]coordinates {
(0.285, 1.6)
(1.5, 1.48)
(1.0, 1.02)
};
\addplot+ [color = black,
draw opacity = 1.0,
line width = 1,
solid,mark = none,forget plot]coordinates {
(0.0, 0.5)
(0.0, 0.0)
};
\addplot+ [color = black,
draw opacity = 1.0,
line width = 1,
solid,mark = none,forget plot]coordinates {
(0.0, 0.0)
(1.0, 1.02)
};
\addplot+ [color = black,
draw opacity = 1.0,
line width = 1,
solid,mark = none,forget plot]coordinates {
(1.0, 1.02)
(0.0, 0.5)
};
\addplot+ [color = black,
draw opacity = 1.0,
line width = 1,
solid,mark = none,forget plot]coordinates {
(2.25, 1.1)
(2.6, 1.1)
};
\addplot+ [color = black,
draw opacity = 1.0,
line width = 1,
solid,mark = none,forget plot]coordinates {
(2.6, 1.1)
(2.0, 3.0)
};
\addplot+ [color = black,
draw opacity = 1.0,
line width = 1,
solid,mark = none,forget plot]coordinates {
(2.0, 3.0)
(2.25, 1.1)
};
\addplot+ [color = black,
draw opacity = 1.0,
line width = 1,
solid,mark = none,forget plot]coordinates {
(0.0, 1.5)
(0.0, 0.5)
};
\addplot+ [color = black,
draw opacity = 1.0,
line width = 1,
solid,mark = none,forget plot]coordinates {
(0.0, 0.5)
(0.285, 1.6)
};
\addplot+ [color = black,
draw opacity = 1.0,
line width = 1,
solid,mark = none,forget plot]coordinates {
(0.285, 1.6)
(0.0, 1.5)
};
\addplot+ [color = black,
draw opacity = 1.0,
line width = 1,
solid,mark = none,forget plot]coordinates {
(1.3, 3.1)
(0.0, 3.0)
};
\addplot+ [color = black,
draw opacity = 1.0,
line width = 1,
solid,mark = none,forget plot]coordinates {
(0.0, 3.0)
(0.285, 1.6)
};
\addplot+ [color = black,
draw opacity = 1.0,
line width = 1,
solid,mark = none,forget plot]coordinates {
(0.285, 1.6)
(1.3, 3.1)
};
\addplot+ [color = black,
draw opacity = 1.0,
line width = 1,
solid,mark = none,forget plot]coordinates {
(1.0, 1.02)
(1.5, 1.48)
};
\addplot+ [color = black,
draw opacity = 1.0,
line width = 1,
solid,mark = none,forget plot]coordinates {
(1.5, 1.48)
(0.285, 1.6)
};
\addplot+ [color = black,
draw opacity = 1.0,
line width = 1,
solid,mark = none,forget plot]coordinates {
(0.285, 1.6)
(1.0, 1.02)
};
\addplot+ [color = black,
draw opacity = 1.0,
line width = 1,
solid,mark = none,forget plot]coordinates {
(0.0, 0.0)
(1.7, 0.2)
};
\addplot+ [color = black,
draw opacity = 1.0,
line width = 1,
solid,mark = none,forget plot]coordinates {
(1.7, 0.2)
(1.0, 1.02)
};
\addplot+ [color = black,
draw opacity = 1.0,
line width = 1,
solid,mark = none,forget plot]coordinates {
(1.0, 1.02)
(0.0, 0.0)
};
\addplot+ [color = black,
draw opacity = 1.0,
line width = 1,
solid,mark = none,forget plot]coordinates {
(0.285, 1.6)
(0.0, 0.5)
};
\addplot+ [color = black,
draw opacity = 1.0,
line width = 1,
solid,mark = none,forget plot]coordinates {
(0.0, 0.5)
(1.0, 1.02)
};
\addplot+ [color = black,
draw opacity = 1.0,
line width = 1,
solid,mark = none,forget plot]coordinates {
(1.0, 1.02)
(0.285, 1.6)
};
\addplot+ [color = black,
draw opacity = 1.0,
line width = 1,
solid,mark = none,forget plot]coordinates {
(0.285, 1.6)
(1.5, 1.48)
};
\addplot+ [color = black,
draw opacity = 1.0,
line width = 1,
solid,mark = none,forget plot]coordinates {
(1.5, 1.48)
(1.3, 3.1)
};
\addplot+ [color = black,
draw opacity = 1.0,
line width = 1,
solid,mark = none,forget plot]coordinates {
(1.3, 3.1)
(0.285, 1.6)
};
\addplot+ [color = black,
draw opacity = 1.0,
line width = 1,
solid,mark = none,forget plot]coordinates {
(0.0, 3.0)
(0.0, 1.5)
};
\addplot+ [color = black,
draw opacity = 1.0,
line width = 1,
solid,mark = none,forget plot]coordinates {
(0.0, 1.5)
(0.285, 1.6)
};
\addplot+ [color = black,
draw opacity = 1.0,
line width = 1,
solid,mark = none,forget plot]coordinates {
(0.285, 1.6)
(0.0, 3.0)
};
\addplot+ [color = black,
draw opacity = 1.0,
line width = 1,
solid,mark = none,forget plot]coordinates {
(3.0, 0.0)
(2.6, 1.1)
};
\addplot+ [color = black,
draw opacity = 1.0,
line width = 1,
solid,mark = none,forget plot]coordinates {
(2.6, 1.1)
(2.25, 1.1)
};
\addplot+ [color = black,
draw opacity = 1.0,
line width = 1,
solid,mark = none,forget plot]coordinates {
(2.25, 1.1)
(3.0, 0.0)
};
\addplot+ [color = black,
draw opacity = 1.0,
line width = 1,
solid,mark = none,forget plot]coordinates {
(2.25, 1.1)
(2.0, 3.0)
};
\addplot+ [color = black,
draw opacity = 1.0,
line width = 1,
solid,mark = none,forget plot]coordinates {
(2.0, 3.0)
(1.5, 1.48)
};
\addplot+ [color = black,
draw opacity = 1.0,
line width = 1,
solid,mark = none,forget plot]coordinates {
(1.5, 1.48)
(2.25, 1.1)
};
\addplot+ [color = black,
draw opacity = 1.0,
line width = 1,
solid,mark = none,forget plot]coordinates {
(1.0, 1.02)
(1.7, 0.2)
};
\addplot+ [color = black,
draw opacity = 1.0,
line width = 1,
solid,mark = none,forget plot]coordinates {
(1.7, 0.2)
(1.5, 1.48)
};
\addplot+ [color = black,
draw opacity = 1.0,
line width = 1,
solid,mark = none,forget plot]coordinates {
(1.5, 1.48)
(1.0, 1.02)
};
\addplot+ [color = black,
draw opacity = 1.0,
line width = 1,
solid,mark = none,forget plot]coordinates {
(1.3, 3.1)
(1.5, 1.48)
};
\addplot+ [color = black,
draw opacity = 1.0,
line width = 1,
solid,mark = none,forget plot]coordinates {
(1.5, 1.48)
(2.0, 3.0)
};
\addplot+ [color = black,
draw opacity = 1.0,
line width = 1,
solid,mark = none,forget plot]coordinates {
(2.0, 3.0)
(1.3, 3.1)
};
\addplot+ [color = black,
draw opacity = 1.0,
line width = 1,
solid,mark = none,forget plot]coordinates {
(1.7, 0.2)
(3.0, 0.0)
};
\addplot+ [color = black,
draw opacity = 1.0,
line width = 1,
solid,mark = none,forget plot]coordinates {
(3.0, 0.0)
(2.25, 1.1)
};
\addplot+ [color = black,
draw opacity = 1.0,
line width = 1,
solid,mark = none,forget plot]coordinates {
(2.25, 1.1)
(1.7, 0.2)
};
\addplot+ [color = black,
draw opacity = 1.0,
line width = 1,
solid,mark = none,forget plot]coordinates {
(1.7, 0.2)
(2.25, 1.1)
};
\addplot+ [color = black,
draw opacity = 1.0,
line width = 1,
solid,mark = none,forget plot]coordinates {
(2.25, 1.1)
(1.5, 1.48)
};
\addplot+ [color = black,
draw opacity = 1.0,
line width = 1,
solid,mark = none,forget plot]coordinates {
(1.5, 1.48)
(1.7, 0.2)
};
\addplot+ [color = black,
draw opacity = 1.0,
line width = 1, ->,
solid,mark = none,forget plot]coordinates {
(1.0, 1.02)
(1.25, 1.25)
};
\addplot+ [color = black,
draw opacity = 1.0,
line width = 1,
solid,mark = none,forget plot]coordinates {
(1.25, 1.25)
(1.5, 1.48)
};
\addplot+ [color = black,
draw opacity = 1.0,
line width = 1, ->,
solid,mark = none,forget plot]coordinates {
(1.5, 1.48)
(0.8925, 1.54)
};
\addplot+ [color = black,
draw opacity = 1.0,
line width = 1,
solid,mark = none,forget plot]coordinates {
(0.8925, 1.54)
(0.285, 1.6)
};
\addplot+ [color = black,
draw opacity = 1.0,
line width = 1, ->,
solid,mark = none,forget plot]coordinates {
(0.285, 1.6)
(0.6425, 1.31)
};
\addplot+ [color = black,
draw opacity = 1.0,
line width = 1,
solid,mark = none,
forget plot]coordinates {
(0.6425, 1.31)
(1.0, 1.02)
};
\addplot+[draw=none, color = black,
draw opacity = 1.0,
line width = 0,
solid,mark = *,
mark size = 2.0,
mark options = {
    color = black, draw opacity = 1.0,
    fill = black, fill opacity = 1.0,
    line width = 1,
    rotate = 0,
    solid
},forget plot] coordinates {
(1.5, 1.48)
};
\addplot+[draw=none, color = black,
draw opacity = 1.0,
line width = 0,
solid,mark = *,
mark size = 2.0,
mark options = {
    color = black, draw opacity = 1.0,
    fill = black, fill opacity = 1.0,
    line width = 1,
    rotate = 0,
    solid
},forget plot] coordinates {
(0.285, 1.6)
};
\addplot+[draw=none, color = black,
draw opacity = 1.0,
line width = 0,
solid,mark = *,
mark size = 2.0,
mark options = {
    color = black, draw opacity = 1.0,
    fill = black, fill opacity = 1.0,
    line width = 1,
    rotate = 0,
    solid
},forget plot] coordinates {
(1.0, 1.02)
};
\node at (axis cs:0.8925, 1.54) [above, color=black] {$e_1$};
\node at (axis cs:1.25, 1.25) [below right, color=black] {$e_2$};
\node at (axis cs:0.6425, 1.31) [below left, color=black] {$e_3$};
\node at (axis cs:1.5, 1.48) [right, color=black] {$v_1$};
\node at (axis cs:0.285, 1.6)[above left, color=black] {$v_2$};
\node at (1.0, 1.02) [below, color=black] {$v_3$};
\end{axis}

\end{tikzpicture}}\quad
\subfloat[Same triangle, with a line field superimposed and values at vertices in red]{
\input{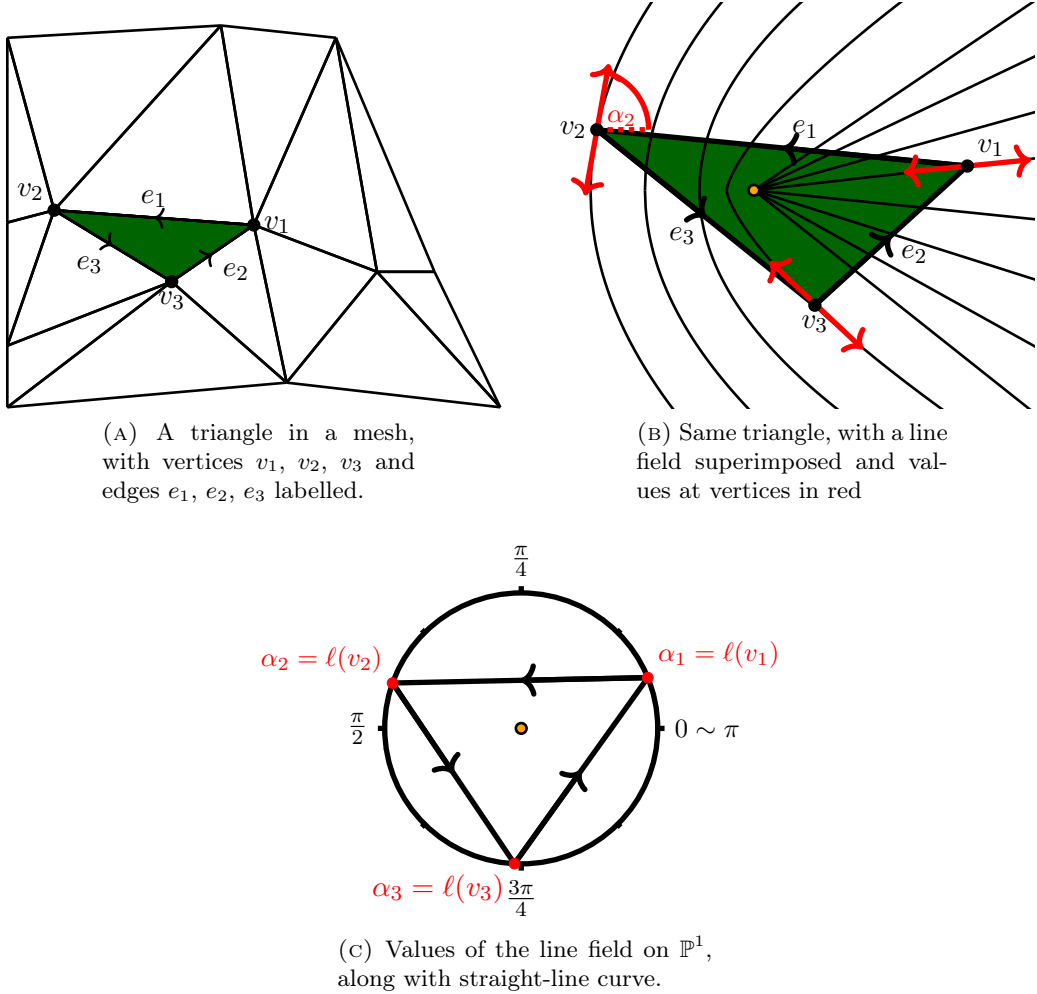}}\\
\subfloat[\label{fig:rp1}Values of the line field on $\mathbb P^1$, along with straight-line curve.]{
    \begin{tikzpicture}

\begin{axis}[
width={0.7\textwidth},
height={0.41\textwidth}, 
axis lines=none,
axis equal image,
xlabel = {}, ylabel = {},
xmin = {-2.3}, xmax = {2.3}, ymin = {-1.5}, ymax = {1.5},
unbounded coords=jump,
]

\draw[line width=2] (axis cs:0.0,0.0) circle [radius=1.0];

\addplot+ [
line width = 2,color = black,
solid,mark = none,
mark size = 2.0,
mark options = {
    color = black, 
    fill = black, 
    line width = 1,
    solid
},forget plot]coordinates {
(0.0, 1.0)
(0.0, 1.05)
};
\addplot+ [
line width = 2,color = black,
solid,mark = none,
mark size = 2.0,
mark options = {
    color = black, 
    fill = black, 
    line width = 1,
    solid
},forget plot]coordinates {
(-1.0, 0.0)
(-1.05, 0.0)
};
\addplot+ [
line width = 2,color = black,
solid,mark = none,
mark size = 2.0,
mark options = {
    color = black, 
    fill = black, 
    line width = 1,
    solid
},forget plot]coordinates {
(0.0, -1.0)
(0.0, -1.05)
};
\addplot+ [
line width = 2,color = black,
solid,mark = none,
mark size = 2.0,
mark options = {
    color = black, 
    fill = black, 
    line width = 1,
    solid
},forget plot]coordinates {
(1.0, 0.0)
(1.05, 0.0)
};
\addplot+ [
line width = 2,color = black,
solid,mark = none,
mark size = 2.0,
mark options = {
    color = black, 
    fill = black, 
    line width = 1,
    solid
},forget plot]coordinates {
(0.7071067811865476, 0.7071067811865475)
(0.728319984622144, 0.7283199846221439)
};
\addplot+ [
line width = 2,color = black,
solid,mark = none,
mark size = 2.0,
mark options = {
    color = black, 
    fill = black, 
    line width = 1,
    solid
},forget plot]coordinates {
(-0.7071067811865475, 0.7071067811865476)
(-0.7283199846221439, 0.728319984622144)
};
\addplot+ [
line width = 2,color = black,
solid,mark = none,
mark size = 2.0,
mark options = {
    color = black, 
    fill = black, 
    line width = 1,
    solid
},forget plot]coordinates {
(-0.7071067811865477, -0.7071067811865475)
(-0.7283199846221441, -0.7283199846221439)
};
\addplot+ [
line width = 2,color = black,
solid,mark = none,
mark size = 2.0,
mark options = {
    color = black, 
    fill = black, 
    line width = 1,
    solid
},forget plot]coordinates {
(0.7071067811865474, -0.7071067811865477)
(0.7283199846221438, -0.7283199846221441)
};

\addplot+ [
line width = 2,color = black,
solid,mark = none,
forget plot]coordinates {
(0.9271735283481612, 0.3746321506897417)
(-0.9422223406686581, 0.3349881501559051)
};
\addplot+ [
line width = 2,color = black,
solid,mark = none,
forget plot]coordinates {
(-0.9422223406686581, 0.3349881501559051)
(-0.049183821914170554, -0.998789743470524)
};
\addplot+ [
line width = 2,color = black,
solid,mark = none,
forget plot]coordinates {
(-0.049183821914170554, -0.998789743470524)
(0.9271735283481612, 0.3746321506897417)
};
\addplot+ [
line width = 0,color = black,
solid,mark = *, mark options = {color=red},
forget plot]coordinates {
(0.9271735283481612, 0.3746321506897417)
(-0.9422223406686581, 0.3349881501559051)
(-0.049183821914170554, -0.998789743470524)
};

\addplot+[draw=none, color = {rgb,1:red,1.00000000;green,0.64705882;blue,0.00000000},
line width = 0,
solid,mark = *,
mark size = 2.0,
mark options = {
    color = black, 
    fill = {rgb,1:red,1.00000000;green,0.64705882;blue,0.00000000}, 
    line width = 1,
    solid
},forget plot] coordinates {
(0.0, 0.0)
};
\addplot+ [
line width = 2,color = black,
solid,mark = none, ->,
forget plot]coordinates {
(0.9271735283481612, 0.3746321506897417)
(-0.007524406160248409, 0.3548101504228234)
};
\addplot+ [
line width = 2,color = black,
solid,mark = none, ->,
forget plot]coordinates {
(-0.9422223406686581, 0.3349881501559051)
(-0.4957030812914143, -0.3319007966573094)
};
\addplot+ [
line width = 2,color = black,
solid,mark = none, ->,
forget plot]coordinates {
(-0.049183821914170554, -0.998789743470524)
(0.43899485321699533, -0.31207879639039116)
};
\node at (axis cs:1.05, 0.0) [right] {\small$0\sim\pi$};
\node at (axis cs:0.0, 1.05) [above] {$\frac{\pi}{4}$};
\node at (axis cs:-1.05, 0.0) [left] {$\frac{\pi}{2}$};
\node at (axis cs:0.0, -1.05) [below] {$\frac{3 \pi}{4}$};
\node at (axis cs:0.9271735283481612, 0.3746321506897417) [above right, color=red] {\small$\alpha_1=\ell(v_1)$};
\node at (axis cs:-0.9422223406686581, 0.3349881501559051) [above left, color=red] {\small$\alpha_2=\ell(v_2)$};
\node at (axis cs:-0.049183821914170554, -0.998789743470524) [below left, color=red] {$\alpha_3=\ell(v_3)$};
\end{axis}

\end{tikzpicture}
    }
\caption{(a) Visualization of an irregular computational domain, a triangular mesh. (b)
The line field $\lf$ and its values $\lf(v_1)$, $\lf(v_2)$, $\lf(v_3)$ around a triangle of
the mesh. (c) The angle (mod $\pi$) representation $\alpha_i = \lf(v_i)$ and a
connecting line of straight line segments. The values $\theta_i$ are (directed) arcs
from $\alpha_i$ to $\alpha_{i+1}$. As the curve goes around the center
halfway, the line-field index is $\frac{1}{2}$, correctly indicating the enclosed wedge-type
singularity.}
\label{fig:triangle}
\end{figure}

From the definition of $\Delta_i$, we know that it changes sign if the direction of
$e_i$ is reversed. This gives an additive property that is consistent with the additive
property of the index.

\begin{lemma}\label{lem:union_index}
Let $P_1,\dots, P_k\in\mathcal{F}$ be $k$ distinct faces so that $R = \bigcup_{i=1}^k P_k$.
Then $\ind_\lf(R) = \sum_{i=1}^k \ind_l P_i$.
\end{lemma}

\Cref{lem:union_index} allows us to ignore faces with vanishing index from all considerations in the following.

The method just described can also be interpreted as follows.
Define the values of $\theta_i$ by taking the canonical metric on $\mathbb P^1$ as
given by the angle between subspaces, and obtain a curve by connecting individual points by the
shortest path in this metric (or equivalently by straight lines in the canonical embedding into $\mathbb R^2$). To compute the index, we then count the number of times
this curve winds around the center of the circle representing $\mathbb{P}^1$, and divide by 2; cf.~the (equivalent) definition of the line field index in \cite[p.~218]{Spivak1999}.

\subsubsection{Step (ii): Combining polygons}
\label{ssec:combination}

Let $\mathcal{F} = \lbrace P_i; ~i \in I\rbrace$ be the set of polygonal faces/grid cells of the mesh enumerated by an index set $I$.
We identify each polygon $P_i \in\mathcal{F}$ with its center of mass $p_i$.
In the following, the distance between faces $P_i$ and $P_j$ is taken as the
distance between the centers of mass $p_i$ and $p_j$, for simplicity.
We wish to detect regions $R$ that are elliptic and $r$-stable unions of polygons. We do so by
finding connected components of an undirected graph $\mathcal{G}$ whose nodes are the
center points $(p_i)_{i \in I}$ and whose faces $P_i$ have non-vanishing index. In
this graph, two nodes $p_i \neq p_j$ are connected if and only if $\lvert p_i - p_j\rvert < r$.
By \cref{lem:union_index}, the index of such a connected component
$\widetilde{\mathcal{G}}$ is given by the sum of the non-vanishing indices
$\sum_{i:~p_i\in\widetilde{\mathcal{G}}}\ind_{\lf}(P_i)$\,.

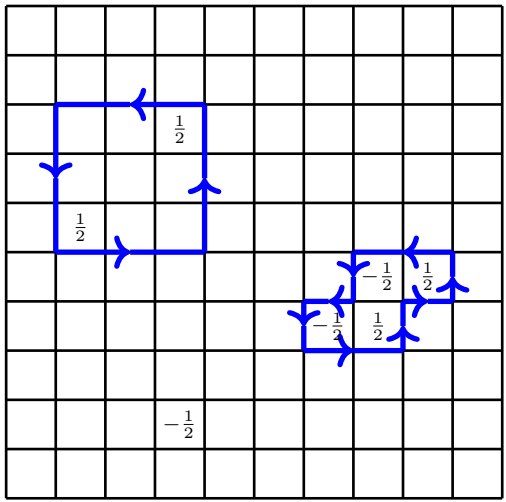
\begin{figure}
\centering
\begin{tikzpicture}[]
\begin{axis}[width = {0.5\textwidth}, height = {0.5\textwidth}, axis lines=none,
xmin = {-0.03}, xmax = {1.03}, ymin = {-0.03}, ymax = {1.03},
xlabel = {}, ylabel = {}, unbounded coords=jump,scaled x ticks = false,
xtick = {},xticklabels = {},xtick align = inside,
grid = minor,
x grid style = {color = black,draw opacity = 1.0,line width = 0.5,solid},
axis x line* = left,x axis line style = {color = black,draw opacity = 1.0,line width = 1,solid},
scaled y ticks = false,
ytick = {0.0,0.1,0.2},yticklabels = {},
y grid style = {color = black,draw opacity = 1.0,line width = 0.5,solid},axis y line* = left,
y axis line style = {color = black,draw opacity = 1.0,line width = 1,solid},
xshift = 0.0mm, yshift = 0.0mm,
]

\foreach \value in {0.0,0.1,...,1.05} {
\addplot+ [color = black, line width = 1,solid, mark = none,forget plot]coordinates {
(\value, 0.0)
(\value, 1.0)
};
\addplot+ [color = black, line width = 1,solid, mark = none,forget plot]coordinates {
(0.0, \value)
(1.0, \value)
};
};

\addplot+ [color = blue, line width = 2, solid, mark = none, ->, forget plot]coordinates {
(0.1, 0.5)
(0.25, 0.5)
};
\addplot+ [color = blue, line width = 2, solid, mark = none, forget plot]coordinates {
(0.25, 0.5)
(0.4, 0.5)
};
\addplot+ [color = blue, line width = 2, solid, mark = none, ->, forget plot]coordinates {
(0.4, 0.5)
(0.4, 0.65)
};
\addplot+ [color = blue, line width = 2, solid, mark = none, forget plot]coordinates {
(0.4, 0.65)
(0.4, 0.8)
};
\addplot+ [color = blue, line width = 2, solid, mark = none, ->, forget plot]coordinates {
(0.4, 0.8)
(0.25, 0.8)
};
\addplot+ [color = blue, line width = 2, solid, mark = none, forget plot]coordinates {
(0.25, 0.8)
(0.1, 0.8)
};
\addplot+ [color = blue, line width = 2, solid, mark = none, ->, forget plot]coordinates {
(0.1, 0.8)
(0.1, 0.65)
};
\addplot+ [color = blue, line width = 2, solid, mark = none, forget plot]coordinates {
(0.1, 0.65)
(0.1, 0.5)
};
\addplot+ [color = blue, line width = 2, solid, mark = none, ->, forget plot]coordinates {
(0.6, 0.3)
(0.7, 0.3)
};
\addplot+ [color = blue, line width = 2, solid, mark = none, forget plot]coordinates {
(0.7, 0.3)
(0.8, 0.3)
};
\addplot+ [color = blue, line width = 2, solid, mark = none, ->, forget plot]coordinates {
(0.8, 0.3)
(0.8, 0.35)
};
\addplot+ [color = blue, line width = 2, solid, mark = none, forget plot]coordinates {
(0.8, 0.35)
(0.8, 0.4)
};
\addplot+ [color = blue, line width = 2, solid, mark = none, ->, forget plot]coordinates {
(0.8, 0.4)
(0.85, 0.4)
};
\addplot+ [color = blue, line width = 2, solid, mark = none, forget plot]coordinates {
(0.85, 0.4)
(0.9, 0.4)
};
\addplot+ [color = blue, line width = 2, solid, mark = none, ->, forget plot]coordinates {
(0.9, 0.4)
(0.9, 0.45)
};
\addplot+ [color = blue, line width = 2, solid, mark = none, forget plot]coordinates {
(0.9, 0.45)
(0.9, 0.5)
};
\addplot+ [color = blue, line width = 2, solid, mark = none, ->, forget plot]coordinates {
(0.9, 0.5)
(0.8, 0.5)
};
\addplot+ [color = blue, line width = 2, solid, mark = none, forget plot]coordinates {
(0.8, 0.5)
(0.7, 0.5)
};
\addplot+ [color = blue, line width = 2, solid, mark = none, ->, forget plot]coordinates {
(0.7, 0.5)
(0.7, 0.45)
};
\addplot+ [color = blue, line width = 2, solid, mark = none, forget plot]coordinates {
(0.7, 0.45)
(0.7, 0.4)
};
\addplot+ [color = blue, line width = 2, solid, mark = none, ->, forget plot]coordinates {
(0.7, 0.4)
(0.65, 0.4)
};
\addplot+ [color = blue, line width = 2, solid, mark = none, forget plot]coordinates {
(0.65, 0.4)
(0.6, 0.4)
};
\addplot+ [color = blue, line width = 2, solid, mark = none, ->, forget plot]coordinates {
(0.6, 0.4)
(0.6, 0.35)
};
\addplot+ [color = blue, line width = 2, solid, mark = none, forget plot]coordinates {
(0.6, 0.35)
(0.6, 0.3)
};
\node at (axis cs:0.15, 0.55) [color=black] {$\scriptstyle \frac{1}{2}$};
\node at (axis cs:0.35, 0.75) [color=black] {$\scriptstyle \frac{1}{2}$};
\node at (axis cs:0.75, 0.35) [color=black] {$\scriptstyle \frac{1}{2}$};
\node at (axis cs:0.85, 0.45) [color=black] {$\scriptstyle \frac{1}{2}$};
\node at (axis cs:0.75, 0.45) [color=black] {$\scriptstyle -\frac{1}{2}$};
\node at (axis cs:0.65, 0.35) [color=black] {$\scriptstyle -\frac{1}{2}$};
\node at (axis cs:0.35, 0.15) [color=black] {$\scriptstyle -\frac{1}{2}$};
\end{axis}

\end{tikzpicture}
\caption{A quadrilateral mesh, with several $r$-stable regions
($r$ equals, e.g., two cell diameters)
containing singular cells. The upper left contour encloses an elliptic region,
whereas the right contour contains 4 cells with non-vanishing indices which sum up to zero.
Below, there is an isolated cell of index $-\frac12$.}
\label{fig:combining}
\end{figure}

Let $\mathcal{K}$ denote the set of such connected components. Any
$K\in\mathcal{K}$ represents a set of faces $R_K \coloneqq \cup_{k \in K} P_k$ whose
index is, according to \cref{lem:union_index} given by the sum of the indices of
$P_k$, $k\in K$. If these faces have $\ind_\lf(K) = 1$, then the corresponding
region $R_K$ is an elliptic region that is (approximately\footnote{As we are working with center points and not exact distances, it may not be fully $r$-stable.})
$r$-stable (cf.~\cref{fig:combining}), provided that the computed index approximates the true index well enough at the chosen coarseness of the polygonal mesh.
For a region $K$, we will call the average over $(p_i)_{i \in K}$ its \emph{center}.

\subsubsection{Step (iii): Additional merge heuristics}\label{sec:mergeheuristics}
As mentioned above, observations in \cite{Karrasch2015} showed that the
procedure in step (ii) may miss large elliptic regions, in which two wedge-type
singularities (with index $\frac12$) are further apart than $r$. Thus, we account for
special elliptic configurations of $r$-stable regions by a range of merge heuristics.
The simplest of these, which we call \texttt{combine\_20}, adds to the list of elliptic regions all wedge pairs
comprising $r$-stable regions that are mutually their nearest neighbors (measured by distance between center points) among the
$r$-stable regions (of nonvanishing index). We have a similar heuristic for
combining 3 wedge-type $r$-stable regions with a trisector (index $-\frac{1}{2}$)
called \texttt{combine\_31}. This seems useful in the OECS case but less so for
other types of vortices. Additionally, we have also implemented a
\texttt{combine\_20\_aggressive} heuristic that combines wedge-pairs under strictly
weaker requirements. More specifically, it does so if (i) one of them is the nearest
neighbor of the other; and (ii) if the rectangle with vertices given by the $r$-stable
region centers does not contain any further $r$-stable region of nonzero index.
This heuristic is based on examination of the singularity configurations occurring in
our turbulence simulation described in \cref{sec:turbulence}. There, we also
compare results from the \texttt{combine\_20} heuristic with those obtained from the
\texttt{combine\_20\_aggressive} heuristic. Further heuristics can be developed and
neatly included in our implementation. Since the resulting regions only serve as
\emph{candidate} regions for the closed orbit computation described in
\cref{sec:closed_orbits}, false-positives at worst add some computational effort.

\subsubsection{Discussion}

Since the above procedure forms one centerpiece of our implementation and
earlier implementations of the same index-theory-based considerations
have deservedly earned some criticism, we would like to discuss some of its
features from a theoretical viewpoint here.
\newpage
First, the procedure described above is a rigorously justified singularity
simplification procedure; cf.~also \cite{Tricoche2001}.
If a singularity contained in a single mesh cell with non-vanishing
index is reasonably isolated, merging its enclosing cell with neighboring cells
results in a curve homotopy of the boundary which does not change the value of the index, but
instead increases the number of ``quadrature points'' in the discretization of its
integral representation; recall \cref{eq:lfindex} in \cref{def:lf_index}. Therefore, it
allows us to compute the index more accurately.
If a singularity is accompanied by a very close second singularity, then their
combined index is computed. This yields either (a) the cancellation of poorly
computed/fake non-zero indices, or (b) the computation of the index of a larger
region enclosing two singularities.
Case (b) sometimes occurs in the center of closed orbits, where two wedges (each
with index $1/2$) or three wedges and a trisector (index $-\frac12$) are nearby and
get combined to a joint singularity with index 1, correctly indicating an elliptic region.
Case (a) often occurs along the observed boundary of vortices, where many
singularities cluster along a line. From a macroscopic perspective, however, the
indices of these singularities turn out to cancel each other out, indicating that
the net topological effect of this cluster is equivalent to the complete absence of
singularities, or the presence of a single singularity.

\begin{figure}
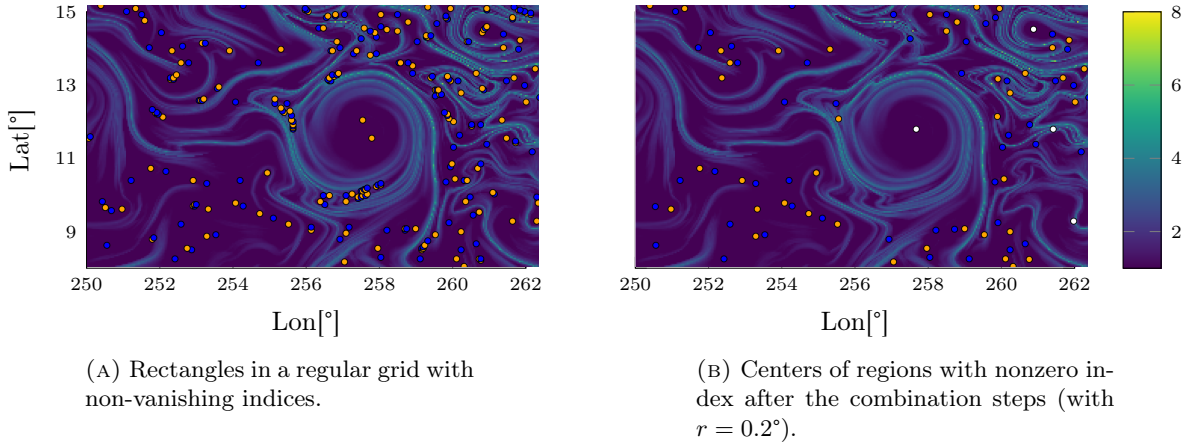

\centering
\subfloat[Rectangles in a regular grid with non-vanishing indices.\label{fig:combi1}]{
\input{sings_before.tex}
}\quad
\subfloat[Centers of regions with nonzero index after the combination steps (with $r=0.2\degree{}$).\label{fig:combi2}]{
\input{sings_after.tex}
}
\caption{Distribution of transport tensor $\mathbf{T}$ singularities and their
types for a $30$-day ocean surface simulation off the coast of Mexico starting on January 7, 2017: orange points
correspond to centers of regions with index $\frac{1}{2}$, blue to index $-\frac{1}{2}$,
and white to index $1$. Background coloring is the DBS field.}
\label{fig:combination}
\end{figure}

These effects are shown in \cref{fig:combination}. In \cref{fig:combi1},
cells with nonzero index on a quadrilateral mesh (only wedges (index $\frac12$, orange) and trisectors (index $-\frac12$,
blue)) without any combination steps taken are shown. One can clearly see the isolated
wedge pair in the center of the figure, and dense singularity clusters aligned along
quasi-one-dimensional strips. In \cref{fig:combi2}, singularities are post-processed
according to the above procedure. Here, (i) the isolated wedge pair is combined to
an elliptic region (index 1, white), (ii) the dense singularity clusters have been
annihilated under combination/index summation, and (iii) a densely clustered 3-wedge-1-trisector
configuration in the upper right corner (caused in this particular case by the use of a
low order interpolation scheme for the velocity field)  has been combined to
another elliptic region.

Finally, compared to the preceding (the computation of the tensor field $\mathbf{T}$
via advection of potentially dense grids of particles) and the subsequent
computational steps (the computation of closed null-geodesics), the index
computations here are negligible in terms of computational effort. This effort consists
of:
\begin{enumerate}
	\item computing pairwise distances between points in the plane, though only distances between a \emph{small} number (depending on the heuristic) of \emph{nearby}
	points are required, which massively reduces the number of possible singularity
	combinations and allows to use a tree structure with \texttt{NearestNeighbors.jl}
	\cite{Carlsson2018} instead of a distance matrix; and
	\item applying some simple logic/filtering on the resulting distance graph.
\end{enumerate}
The possibility to obtain information on where closed
null-geodesics might be located allows for the restriction to a small (in comparison to the total number of mesh vertices)
number of local domains, which facilitates good scaling behavior when dealing with challenging flow problems on large domains.

\subsection{Restriction to local domains}
\label{sec:localization}

One crucial feature of the index-based computation of geodesic vortices is the
possibility to localize the subsequent closed-orbit computation to regions of a
physically reasonable size. This distinguishes our implementation from that
developed in \cite{Serra2017}. One immediate and significant effect of data
localization is that interpolants do not have to carry global information, allowing
for a better utilization of computer memory (and the memory hierarchy) in the subsequent steps.
This is especially true when working with several processes, as the amount of data
needing to be sent to individual processes is significantly reduced.

Another positive effect of localization is that integral curves which leave
the localization box trigger an error in the ODE integrator and are hence no longer
followed. Finding a reasonable criterion for automated decision-making when to give
up on an integral curve was a rather challenging issue in previous implementations.

The side length $2R$ of the localization box is a parameter that must be supplied and
that bounds the size of geodesic vortices that can be found. This parameter is used
by our implementation to determine default values for other parameters as well.
These include the required return distance for an integral curve to be considered as
``closed'' and error tolerances of the ODE solver. Each candidate region identified in the
previous step can be processed with only local information about the tensor
eigenvector $\xi_1,\xi_2$ and eigenvalue $\lambda_1,\lambda_2$ fields.

\subsection{Computation of closed orbits}\label{sec:closed_orbits}

Let $K$ be an elliptic region identified by the procedure described in \cref{sec:elliptic}.
That procedure returns a point $p_K$ (typically lying in the convex hull of $K$) which we view
as the potential vortex center. To find closed orbits of $\eta^{\pm}_\lambda$ near $K$,
we employ a shooting method and place a Poincaré section of length $R$ from $p_K$ eastwards to $p_K + (R,0)$.
In order to apply the shooting method, i.e., to numerically\footnote{We use the
\texttt{DifferentialEquations.jl} package \cite{Rackauckas2017}.} compute integral curves,
we need to turn the line fields into local vector fields.

\subsubsection{Orientation of the line field}

At each point $x$ in the domain, the line field $\eta^\pm_\lambda(x)$ is identified with a vector
$v^\pm_\lambda(x)$ so that $\spn\lbrace v_\lambda^\pm(x)\rbrace  = \eta^\pm_\lambda(x)$
and $|v_\lambda^\pm(x)| = 1$. Clearly, integral curves of $v^\pm_\lambda$ are integral curves
of $\eta^\pm_\lambda$. As for the vector field representation, we choose an orientation such
that $\xi_1$ is spiraling anti-clockwise around $p_K$, and $\xi_2$ is pointing
away from $p_K$. This is achieved by setting
\begin{align*}
\tilde \xi_1(x) &\coloneqq \sign\left(\xi_1(x) \cdot \begin{pmatrix} 0 & -1 \\ 1 & 0 \end{pmatrix}(x - p_K)\right)\xi_1(x)\,, &
\tilde \xi_2(x) &\coloneqq \sign\left(\xi_2(x) \cdot (x - p_K)\right)\xi_2(x)\,,
\end{align*}
and then
\begin{align}\label{vformula}
v_\lambda^\pm(x) = \sqrt{\frac{\lambda_2 - \lambda }{\lambda_2 - \lambda_1}}\,\tilde{\xi}_1(x) \pm \sqrt{\frac{\lambda - \lambda_1}{\lambda_2 - \lambda_1}}\,\tilde{\xi}_2(x)\,.
\end{align}
The local vector field can be interpreted as a \emph{rotated vector field}; cf.~\cite{Duff1953}.
With the previous orientation, increasing $\lambda$ turns $v_{\lambda}$ to the right (for $v^+_\lambda$)
and to the left (for $v^-_\lambda$).

We calculate $v_\lambda^\pm$ at grid points and then interpolate. On a quadrilateral grid,
this is done by bilinear interpolation followed by a slight rescaling to ensure that the interpolated
value $v(x)$ has unit length. In cases where $v(x) \cdot v(y) < 0$ for adjacent grid points $x,y$
the interpolated vector field will have sharp kinks, suggesting that the orientation for the vector
field may have been chosen incorrectly. As a post-processing step, we check that all
obtained closed orbits do not lie in such mesh cells. Moreover, \cref{vformula} is undefined when
$\lambda \notin [\lambda_1,\lambda_2]$. In our implementation, we nevertheless generate a
vector field at such points, and reject closed orbits a posteriori if $v_\lambda^\pm$
(and likewise also $\eta_\lambda^\pm$) has points in a cell in which
$v_\lambda^\pm$ is undefined at some vertex. We also reject closed orbits in the
unlikely case that they do not enclose the center point $p_K$.

\subsubsection{The shooting method}

After these preparations, the general approach is the following:
Let $\Ret^\pm\colon [\lambda_{\min},\lambda_{\max}] \times [0,R] \rightarrow \R$
denote the Poincaré return distance function for $v^\pm_\lambda$. This is the
(signed) distance from an orbit of $v^\pm_\lambda$ starting at $p_K + (0,s)$ to its
first return point on the Poincaré section. The function $\Ret$ is undefined at points $s_0$
whose orbit does not return, and in such cases, we assign $\Ret^\pm(s_0)\coloneqq\infty$.
Geodesic vortices are given by the zeroes of $\Ret^\pm$.

Abstractly speaking, the problem is to find roots of $\Ret$ over the square
$[\lambda_{\min},\lambda_{\max}] \times [0,R]$. Previous implementations for computing
geodesic vortices \cite{Onu2015,Karrasch2015,Hadjighasem2016,Serra2017} iterate over a fixed range of values
$\lambda \in [\lambda_{\min},\lambda_{\max}]$ and aim to find zeros of $s\mapsto\Ret^\pm(\lambda,s)$.

\begin{figure}
\centering
\subfloat[The set of points on the Poincaré section for which a given closing parameter $\lambda$ (red lines) corresponds to a closed orbit is difficult to determine for some values of $\lambda$ \label{fig:x_vs_lambda}]{\begin{tikzpicture}[]
\begin{axis}[height = {0.3\textheight}, width = {0.7\textwidth},
xmin = {0.0}, xmax = {1.0}, ymin = {1.097216796875}, ymax = {1.35}, 
xlabel = {distance to vortex center $[\deg]$}, ylabel = {closing parameter},
unbounded coords=jump,
scaled x ticks = false, xtick = {0,.0,0.2,0.4,0.6,0.8,1.0},
xtick align = inside, xticklabel style = {font = \small},
scaled y ticks = false, ytick = {1.1,1.2,1.3,1.4},
ytick align = inside, yticklabel style = {font = \small},
xmajorgrids = true, ymajorgrids = true,
x grid style = {color=black, line width = 0.5, draw opacity = 0.1, solid},
y grid style = {color=black, line width = 0.5, draw opacity = 0.1, solid},
xshift = 0.0mm, yshift = 0.0mm,
legend style = {color = black, draw opacity = 1.0, line width = 1, solid, fill = white, font = {\small}}
]

\addplot+ [color = {rgb,1:red,0.00000000;green,0.60560316;blue,0.97868012},
draw opacity = 1.0,
line width = 2,
solid,mark = none,
mark size = 2.0,
mark options = {
    color = {rgb,1:red,0.00000000;green,0.00000000;blue,0.00000000}, draw opacity = 1.0,
    fill = {rgb,1:red,0.00000000;green,0.60560316;blue,0.97868012}, fill opacity = 1.0,
    line width = 1,
    rotate = 0,
    solid
},forget plot]coordinates {
(0.0448978609754338, 1.1287841796875)
(0.04988651219492635, 1.136474609375)
(0.05487516341441889, 1.140869140625)
(0.05986381463391144, 1.1427001953125)
(0.06485246585340398, 1.14324951171875)
(0.06984111707289697, 1.14251708984375)
(0.07482976829238952, 1.140869140625)
(0.07981841951188207, 1.140869140625)
(0.08480707073137461, 1.1397705078125)
(0.08979572195086716, 1.138671875)
(0.09478437317036015, 1.137939453125)
(0.0997730243898527, 1.136474609375)
(0.10476167560934524, 1.1357421875)
(0.10975032682883779, 1.135009765625)
(0.11473897804833033, 1.133544921875)
(0.11972762926782332, 1.132080078125)
(0.12471628048731587, 1.130615234375)
(0.1297049317068084, 1.129150390625)
(0.13469358292630096, 1.127685546875)
(0.1396822341457935, 1.126220703125)
(0.1446708853652865, 1.124755859375)
(0.14965953658477904, 1.1240234375)
(0.1546481878042716, 1.12255859375)
(0.15963683902376413, 1.12109375)
(0.16462549024325668, 1.11962890625)
(0.16961414146274967, 1.1181640625)
(0.1746027926822422, 1.11669921875)
(0.17959144390173476, 1.115234375)
(0.1845800951212273, 1.115234375)
(0.18956874634071985, 1.11376953125)
(0.19455739756021284, 1.1123046875)
(0.1995460487797054, 1.1123046875)
(0.20453469999919793, 1.11083984375)
(0.20952335121869048, 1.11083984375)
(0.21451200243818302, 1.109375)
(0.21950065365767601, 1.109375)
(0.22448930487716856, 1.10791015625)
(0.2294779560966611, 1.10791015625)
(0.23446660731615365, 1.10791015625)
(0.2394552585356462, 1.10791015625)
(0.2444439097551392, 1.1064453125)
(0.24943256097463173, 1.1064453125)
(0.2544212121941243, 1.1064453125)
(0.2594098634136168, 1.1064453125)
(0.26439851463310937, 1.1064453125)
(0.26938716585260236, 1.1064453125)
(0.2743758170720949, 1.1064453125)
(0.27936446829158745, 1.1064453125)
(0.28435311951108, 1.10791015625)
(0.28934177073057255, 1.10791015625)
(0.2943304219500651, 1.10791015625)
(0.2993190731695581, 1.109375)
(0.3043077243890506, 1.109375)
(0.3092963756085432, 1.109375)
(0.3142850268280357, 1.109375)
(0.31927367804752826, 1.11083984375)
(0.32426232926702125, 1.11083984375)
(0.3292509804865138, 1.11083984375)
(0.33423963170600635, 1.1123046875)
(0.3392282829254989, 1.1123046875)
(0.34421693414499144, 1.115234375)
(0.3492055853644844, 1.115234375)
(0.354194236583977, 1.1181640625)
(0.3591828878034695, 1.11962890625)
(0.36417153902296207, 1.12109375)
(0.3691601902424546, 1.1240234375)
(0.3741488414619476, 1.12548828125)
(0.37913749268144015, 1.126953125)
(0.3841261439009327, 1.1298828125)
(0.38911479512042524, 1.1328125)
(0.3941034463399178, 1.13427734375)
(0.3990920975594108, 1.1357421875)
(0.4040807487789033, 1.138671875)
(0.40906939999839587, 1.14306640625)
(0.4140580512178884, 1.1474609375)
(0.41904670243738096, 1.150390625)
(0.42403535365687395, 1.1533203125)
(0.4290240048763665, 1.15625)
(0.43401265609585904, 1.16064453125)
(0.4390013073153516, 1.162109375)
(0.44398995853484413, 1.1650390625)
(0.4489786097543371, 1.16796875)
(0.45396726097382967, 1.173828125)
(0.4589559121933222, 1.1767578125)
(0.46394456341281476, 1.1826171875)
(0.4689332146323073, 1.185546875)
(0.4739218658518003, 1.1884765625)
(0.47891051707129284, 1.18994140625)
(0.4838991682907854, 1.1943359375)
(0.48888781951027793, 1.19580078125)
(0.4938764707297705, 1.197265625)
(0.49886512194926347, 1.19873046875)
(0.503853773168756, 1.2001953125)
(0.5088424243882486, 1.203125)
(0.5138310756077411, 1.20458984375)
(0.5188197268272337, 1.2060546875)
(0.5238083780467266, 1.20458984375)
(0.5287970292662192, 1.20458984375)
(0.5337856804857117, 1.20458984375)
(0.5387743317052043, 1.20458984375)
(0.5437629829246968, 1.20458984375)
(0.5487516341441898, 1.20458984375)
(0.5537402853636824, 1.20458984375)
(0.5587289365831749, 1.20458984375)
(0.5637175878026675, 1.20458984375)
(0.56870623902216, 1.20458984375)
(0.573694890241653, 1.2060546875)
(0.5786835414611455, 1.20751953125)
(0.5836721926806381, 1.208984375)
(0.5886608439001306, 1.208984375)
(0.5936494951196232, 1.208984375)
(0.5986381463391162, 1.208984375)
(0.6036267975586087, 1.20751953125)
(0.6086154487781013, 1.20751953125)
(0.6136040999975938, 1.2060546875)
(0.6185927512170863, 1.2060546875)
(0.6235814024365789, 1.2060546875)
(0.6285700536560719, 1.2060546875)
(0.6335587048755644, 1.2060546875)
(0.638547356095057, 1.2060546875)
(0.6435360073145495, 1.2060546875)
(0.6485246585340421, 1.2060546875)
(0.653513309753535, 1.2060546875)
(0.6585019609730276, 1.2060546875)
(0.6634906121925201, 1.20751953125)
(0.6684792634120127, 1.20751953125)
(0.6734679146315052, 1.20751953125)
(0.6784565658509982, 1.2060546875)
(0.6834452170704908, 1.2060546875)
(0.6884338682899833, 1.20751953125)
(0.6934225195094759, 1.208984375)
(0.6984111707289684, 1.21044921875)
(0.7033998219484614, 1.21337890625)
(0.708388473167954, 1.21484375)
(0.7133771243874465, 1.21484375)
(0.718365775606939, 1.21484375)
(0.7233544268264316, 1.21337890625)
(0.7283430780459246, 1.208984375)
(0.7333317292654171, 1.203125)
(0.7383203804849097, 1.197265625)
(0.7433090317044022, 1.19140625)
(0.7482976829238948, 1.1884765625)
(0.7532863341433877, 1.185546875)
(0.7582749853628803, 1.18408203125)
(0.7632636365823728, 1.185546875)
(0.7682522878018654, 1.185546875)
(0.7732409390213579, 1.185546875)
(0.7782295902408509, 1.18408203125)
(0.7832182414603435, 1.1826171875)
(0.788206892679836, 1.1826171875)
(0.7931955438993286, 1.1796875)
(0.7981841951188211, 1.1796875)
(0.8031728463383141, 1.1826171875)
(0.8081614975578066, 1.18408203125)
(0.8131501487772992, 1.185546875)
(0.8181387999967917, 1.1884765625)
(0.8231274512162843, 1.19140625)
(0.8281161024357773, 1.19287109375)
(0.8331047536552698, 1.19287109375)
(0.8380934048747624, 1.19140625)
(0.8430820560942549, 1.19140625)
(0.8480707073137475, 1.1943359375)
(0.8530593585332404, 1.19580078125)
(0.858048009752733, 1.197265625)
(0.8630366609722255, 1.197265625)
(0.8680253121917181, 1.2001953125)
(0.8730139634112106, 1.203125)
(0.8780026146307036, 1.208984375)
(0.8829912658501962, 1.2119140625)
(0.8879799170696887, 1.2119140625)
(0.8929685682891813, 1.21630859375)
(0.8979572195086738, 1.220703125)
(0.9029458707281668, 1.2236328125)
(0.9079345219476593, 1.2265625)
(0.9129231731671519, 1.2294921875)
(0.9179118243866444, 1.2353515625)
(0.922900475606137, 1.244140625)
(0.92788912682563, 1.255859375)
(0.9328777780451225, 1.267578125)
(0.937866429264615, 1.2822265625)
(0.9428550804841076, 1.2998046875)
(0.9478437317036001, 1.326171875)
(0.9877529414595414, 1.4140625)
(0.992741592679034, 1.4140625)
};

\addplot+ [color = red, draw opacity = 1.0, line width = 2, solid, mark = none, forget plot]
coordinates {
(0.0, 1.15)
(1.0, 1.15)
};

\addplot+ [color = red, draw opacity = 1.0, line width = 2, solid, mark = none, forget plot]
coordinates {
(0.0, 1.17)
(1.0, 1.17)
};

\addplot+ [color = red, draw opacity = 1.0, line width = 2, solid, mark = none, forget plot]
coordinates {
(0.0, 1.19)
(1.0, 1.19)
};

\addplot+ [color = red, draw opacity = 1.0, line width = 2, solid, mark = none, forget plot]
coordinates {
(0.0, 1.21)
(1.0, 1.21)
};

\addplot+ [color = red, draw opacity = 1.0, line width = 2, solid, mark = none, forget plot]
coordinates {
(0.0, 1.23)
(1.0, 1.23)
};
\end{axis}

\end{tikzpicture}}\\
\subfloat[The return distance function (with fixed parameter value $\lambda$),
whose zeroes correspond to seeding points for closed orbits, may have tangencies with the zero-level set, and, as a consequence, the root finding problem is ill-posed.\label{fig:varying_seeds}]{\input{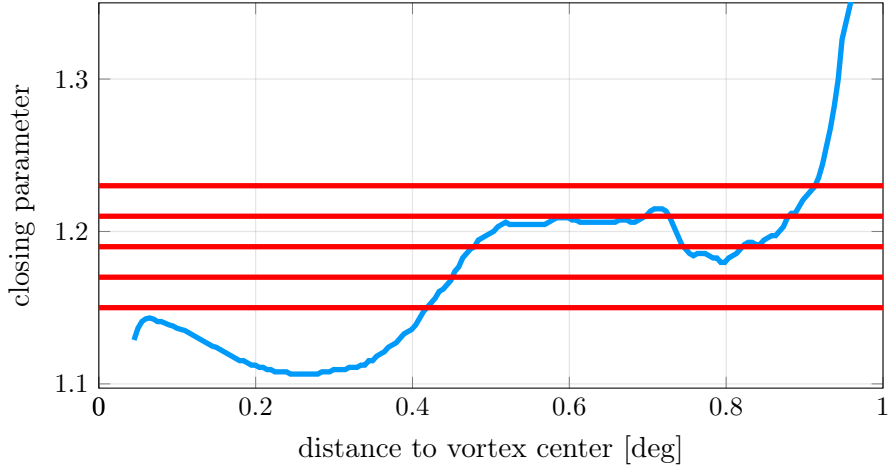}
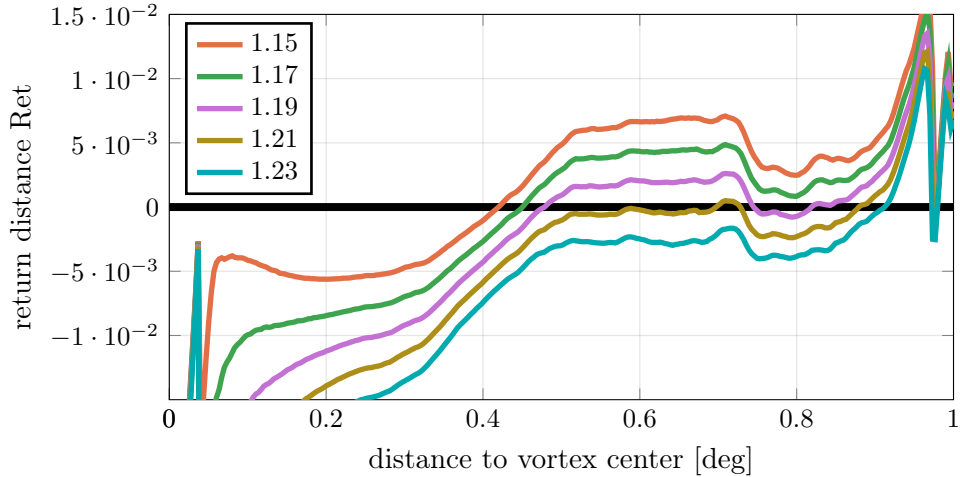}
\caption{Problematic behavior of $\Ret^\pm(\lambda,\cdot)$ with fixed $\lambda$.}
\label{fig:zerocrossings}
\end{figure}

This has the disadvantage of often dealing with poorly conditioned problems.
\Cref{fig:zerocrossings} shows an example of the behavior of $\Ret^\pm(\lambda,\cdot)$
for various values of $\lambda$. In \cref{fig:x_vs_lambda}, we plot the parameter value
$\lambda^*(s)$, for which integral curves close, over the initial condition $(0,s)$ along the Poincaré
section. At local extrema, we observe a jump in the number of closed orbits for fixed
$\lambda$. In \cref{fig:varying_seeds}, we look at the same situation from a different
angle. Here, we plot the return distance function $\Ret$ again over the initial conditions.
The bifurcations in \cref{fig:x_vs_lambda} correspond to tangencies of the curve with the
zero level set $\Ret=0$ (black) in \cref{fig:varying_seeds}.

\begin{figure}
\centering
\subfloat[Return distance function $\Ret$ for some fixed initial condition and varying parameter $\lambda$.]{\begin{tikzpicture}[]
\begin{axis}[scale only axis, height=0.33\textwidth, width = {0.3\textwidth},
xmin = {0}, xmax = {4}, ymin = {-0.19098805320838486}, ymax = {0.1505},
xlabel = {$\lambda$}, ylabel = {return distance $\Ret$}, 
unbounded coords=jump,
scaled x ticks = false, xtick = {0.0,1.0,2.0,3.0,4.0},
xticklabels = {$0$,$1$,$2$,$3$,$4$}, xtick align = inside, xticklabel style = {font = \small},
scaled y ticks = false, ytick = {-0.15000000000000002,-0.1,-0.05,0.0,0.05,0.1},
yticklabels = {$-0.15$,$-0.10$,$-0.05$,$0.00$,$0.05$,$0.10$}, ytick align = inside,
yticklabel style = {font = \small},
xmajorgrids = true, ymajorgrids = true,
x grid style = {color=black, line width = 0.5, draw opacity = 0.1, solid},
y grid style = {color=black, line width = 0.5, draw opacity = 0.1, solid},
xshift = 0.0mm, yshift = 0.0mm
]

\addplot+ [color = {rgb,1:red,0.00000000;green,0.60560316;blue,0.97868012},
draw opacity = 1.0, line width = 3, solid, mark = none, mark size = 2.0,
mark options = {color = {rgb,1:red,0.00000000;green,0.00000000;blue,0.00000000}, draw opacity = 1.0, fill = {rgb,1:red,0.00000000;green,0.60560316;blue,0.97868012}, fill opacity = 1.0, line width = 1, rotate = 0, solid},forget plot]
coordinates {
(0.0, -0.1813939907355122)
(0.04040404040404041, -0.1813939907355122)
(0.08080808080808081, -0.1813939907355122)
(0.12121212121212122, -0.18130174789782671)
(0.16161616161616163, -0.17498283901466216)
(0.20202020202020202, -0.16847925049324886)
(0.24242424242424243, -0.15998540419234164)
(0.2828282828282828, -0.1461181004854586)
(0.32323232323232326, -0.13209764180623162)
(0.36363636363636365, -0.11290001639313285)
(0.40404040404040403, -0.0997518719413315)
(0.4444444444444444, -0.08967774185712196)
(0.48484848484848486, -0.08117756762504946)
(0.5252525252525253, -0.07385576615321066)
(0.5656565656565656, -0.06736115071358251)
(0.6060606060606061, -0.06130955955215356)
(0.6464646464646465, -0.05562900201743792)
(0.6868686868686869, -0.05036706948014702)
(0.7272727272727273, -0.04544592362900035)
(0.7676767676767676, -0.04088317308497391)
(0.8080808080808081, -0.036539422298241586)
(0.8484848484848485, -0.032426847151844296)
(0.8888888888888888, -0.02853470134233671)
(0.9292929292929293, -0.02478694227176037)
(0.9696969696969697, -0.021038574598049475)
(1.0101010101010102, -0.0172242533007827)
(1.0505050505050506, -0.013394809445280043)
(1.0909090909090908, -0.009530119681159022)
(1.1313131313131313, -0.005694901823123821)
(1.1717171717171717, -0.0019286413694246107)
(1.2121212121212122, 0.001796398152213552)
(1.2525252525252526, 0.005364078607672962)
(1.292929292929293, 0.008869065628168471)
(1.3333333333333333, 0.012219533356886636)
(1.3737373737373737, 0.015452182275085047)
(1.4141414141414141, 0.018468832253770717)
(1.4545454545454546, 0.02135496405456605)
(1.494949494949495, 0.02415069891606514)
(1.5353535353535352, 0.026833067143741562)
(1.5757575757575757, 0.02948448184435426)
(1.6161616161616161, 0.03199815087293967)
(1.6565656565656566, 0.03445373433663024)
(1.696969696969697, 0.03680845858180026)
(1.7373737373737375, 0.039100066832639335)
(1.7777777777777777, 0.041298773979186354)
(1.8181818181818181, 0.043458861838319596)
(1.8585858585858586, 0.04556699735174252)
(1.898989898989899, 0.047663438832294514)
(1.9393939393939394, 0.049771909241900225)
(1.97979797979798, 0.05178048555582615)
(2.0202020202020203, 0.05378495863655308)
(2.0606060606060606, 0.055813059421820466)
(2.101010101010101, 0.057827061926588375)
(2.1414141414141414, 0.05976188164473939)
(2.1818181818181817, 0.06173768667014645)
(2.2222222222222223, 0.06368815648537085)
(2.2626262626262625, 0.06566176152783454)
(2.303030303030303, 0.0676373569141111)
(2.3434343434343434, 0.06961081591805085)
(2.3838383838383836, 0.07154611026429158)
(2.4242424242424243, 0.07352656983300454)
(2.4646464646464645, 0.07543771536881794)
(2.505050505050505, 0.07736414236352074)
(2.5454545454545454, 0.07929473171088519)
(2.585858585858586, 0.08122138806970369)
(2.6262626262626263, 0.08318524198288557)
(2.6666666666666665, 0.0851326405434718)
(2.707070707070707, 0.08710880225669504)
(2.7474747474747474, 0.08917659757935947)
(2.787878787878788, 0.09117399732096132)
(2.8282828282828283, 0.09328662689559142)
(2.8686868686868685, 0.09540420882769363)
(2.909090909090909, 0.09749011721006262)
(2.9494949494949494, 0.09957686617737904)
(2.98989898989899, 0.10158227089797833)
(3.0303030303030303, 0.1035589520632656)
(3.0707070707070705, 0.10547053467924883)
(3.111111111111111, 0.10728205819750603)
(3.1515151515151514, 0.10906672774977455)
(3.191919191919192, 0.11076766270550209)
(3.2323232323232323, 0.1124339538690764)
(3.272727272727273, 0.11403331225469726)
(3.313131313131313, 0.11561986946931668)
(3.3535353535353534, 0.11710606982602778)
(3.393939393939394, 0.11856496940419436)
(3.4343434343434343, 0.12000871918256273)
(3.474747474747475, 0.12139906045425342)
(3.515151515151515, 0.12277546930504979)
(3.5555555555555554, 0.1241446814019116)
(3.595959595959596, 0.1254859377460642)
(3.6363636363636362, 0.1269028480304888)
(3.676767676767677, 0.1282303014215289)
(3.717171717171717, 0.1295915556787426)
(3.757575757575758, 0.13086147242402113)
(3.797979797979798, 0.132150407977357)
(3.8383838383838382, 0.13345902852703073)
(3.878787878787879, 0.1346998088681315)
(3.919191919191919, 0.13596770471974162)
(3.95959595959596, 0.13721036344283943)
(4.0, 0.13840809169357549)
};

\addplot+ [color = {rgb,1:red,1.00000000;green,0.00000000;blue,0.00000000},
draw opacity = 1.0, line width = 2, solid, mark=none, forget plot]
coordinates {
(0.0, 0.0)
(4.0, 0.0)
};
\end{axis}

\end{tikzpicture}}\quad
\subfloat[Yellow lines show integral curves for $\lambda=0,1,2,3,4$. Pink lines are with $\lambda=1$ and do not return.\label{fig:lambda}]{\input{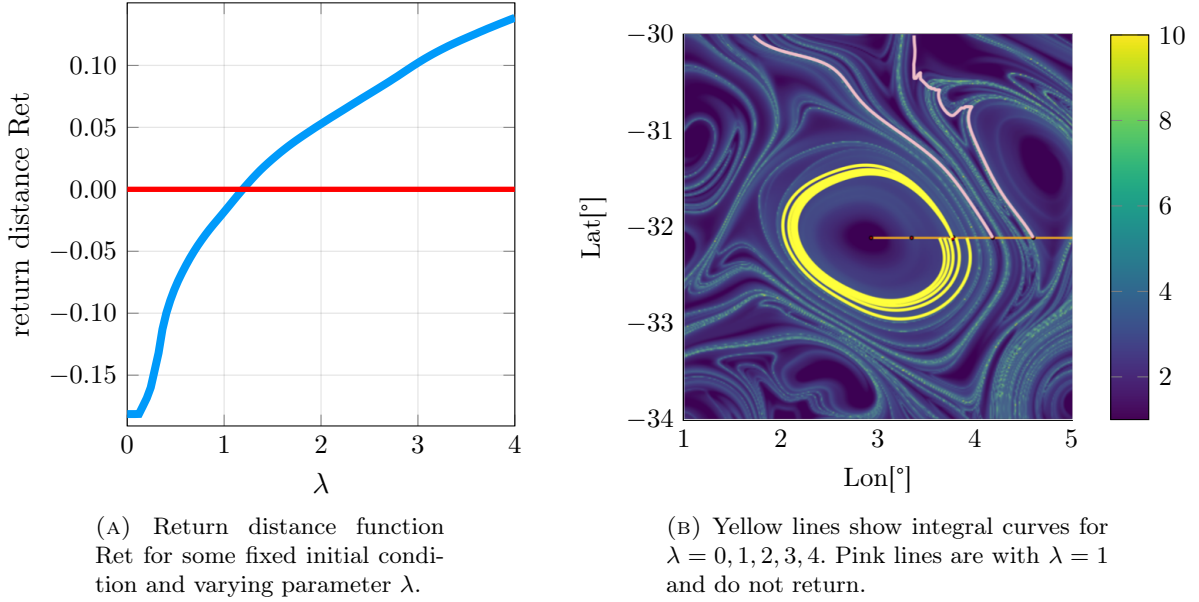}}
\caption{Behavior of the function $\mbox{Ret}^\pm(\cdot, s)$}
\label{fig:prd_vs_lambda}
\end{figure}

\begin{figure}
\centering
\input{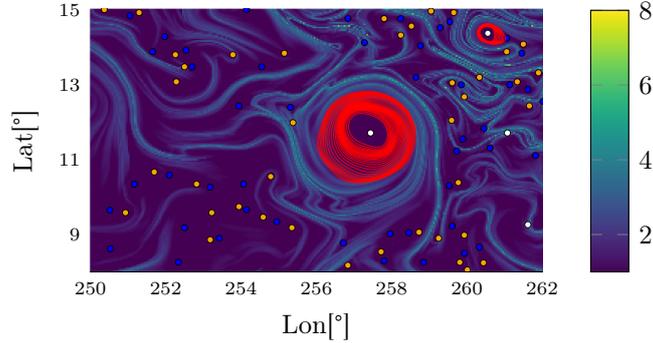}
\caption{\Cref{fig:combi2} overlaid with the computed closed orbits.}
\label{fig:closed_orbits}
\end{figure}

We therefore employ a dual approach instead and look for roots of $\Ret^\pm(\cdot,s)$
for a fixed range of values $s \in [0,R]$, corresponding to a range of fixed initial
conditions along the Poincaré section. This has the following advantages:
\begin{enumerate}
\item The function $\Ret^\pm(\cdot,s)$ is---as long as the corresponding integral curves go through an annular region around the center singularities in which $v_\lambda^\pm$ is a continuous vector field (recall that the underlying vector field is rotating
under parameter variation)---monotone (see \cref{fig:prd_vs_lambda}), greatly improving the condition of the problem.
\item If we only want to find outermost geodesic vortex, we can start with a large $s$, i.e., at the very right of
the Poincaré section, and decrease $s$ until the first closed orbit is found.
\item Closed orbits that are found tend to be uniformly spatially distributed; see \cref{fig:closed_orbits}.
\end{enumerate}

\section{Applications with \texttt{CoherentStructures.jl}}
\label{sec:applications}

Both test cases described below were run on a workstation\footnote{Intel(R) Xeon(R) CPU E3-1235 @ 3.20GHz, 16GB RAM} with 4 cores with a Linux (Fedora 27) operating system, except for the computations used to produce \Cref{fig:vortex_density}, which was run on a 32-core compute server.

\subsection{2D turbulence with varying initial time as parameter}
\label{sec:turbulence}

We calculate material barriers of an incompressible turbulent velocity
field over a series of time windows $[t, t+5]$ for $700$ equally
spaced values of $t$ in $[0,70]$. This is done in order to test
our implementation on a large number of velocity fields without the option of
manually adjusting parameters from one time window to the next.

For the convenience of the reader, we provide complete code---here and as a
Jupyter notebook in the supplementary material---to reproduce our simulation.
Unfortunately, the exact velocity field obtained changes from run to run, but
the results should remain qualitatively the same.
The velocity field is generated with the help of the packages \texttt{FourierFlows.jl} \cite{Wagner2019}
and \texttt{GeophysicalFlows.jl} \cite{Constantinou2019}.
\newpage
\subsubsection{Generating a turbulent velocity field}
We begin by importing the relevant packages and by setting up the computational domain.

\begin{minipage}{\textwidth}
\begin{lstlisting}
using FourierFlows, GeophysicalFlows, GeophysicalFlows.TwoDNavierStokes
mygrid = TwoDGrid(256, 2π)
x, y = gridpoints(mygrid)
xs = ys = range(-π, stop=π, length=257)[1:end-1]
\end{lstlisting}
\end{minipage}

To avoid decay of the flow we employ stochastic forcing.
The code below is modified from the example given in the \texttt{GeophysicalFlows.jl} documentation.

\begin{minipage}{\textwidth}
\begin{lstlisting}
ε = 0.001 # Energy injection rate
kf, dkf = 6, 2.0 # Waveband where to inject energy
Kr = [mygrid.kr[i] for i=1:mygrid.nkr, j=1:mygrid.nl]
force2k = @. exp(-(sqrt(mygrid.Krsq) - kf)^2 / (2 * dkf^2))
force2k[(mygrid.Krsq .< 2.0^2) .| (mygrid.Krsq .> 20.0^2) .| (Kr .< 1) ] .= 0
ε0 = FourierFlows.parsevalsum(force2k .* mygrid.invKrsq/2.0,
                                mygrid) / (mygrid.Lx*mygrid.Ly)
force2k .= ε/ε0 * force2k
function calcF!(Fh, sol, t, cl, args...)
    eta = exp.(2π * im * rand(Float64, size(sol))) / sqrt(cl.dt)
    eta[1,1] = 0
    @. Fh = eta * sqrt(force2k)
    nothing
end
\end{lstlisting}
\end{minipage}

We now setup the remaining parameters used in the simulation. We numerically solve the vorticity (transport) equation
\[
\partial_t \zeta = - u\cdot \nabla \zeta  -\nu\zeta + f.
\]
Here $u(x,y) = (u_1(x,y),u_2(x,y))^T$ is the (incompressible) velocity field, and $\zeta = \partial_x u_2 - \partial_y u_1$ is its vorticity.
The parameter  $\nu$ is set to $10^{-2}$ and is the coefficient of the drag term, $f$ represents the forcing (see also the \texttt{FourierFlows.jl} package and its documentation \cite{Wagner2019}, and \cite{Constantinou2015}).

\begin{minipage}{\textwidth}
\begin{lstlisting}
prob = TwoDNavierStokes.Problem(nx=256, Lx=2π, ν=1e-2, nν=0, dt=1e-2,
    stepper="FilteredRK4", calcF=calcF!, stochastic=true)
TwoDNavierStokes.set_zeta!(prob, GeophysicalFlows.peakedisotropicspectrum(mygrid,2,0.5))

using Distributed
addprocs()
using SharedArrays
# we use these variables to store the result
us = SharedArray{Float64}(256, 256, 400)
vs = SharedArray{Float64}(256, 256, 400)
zs = SharedArray{Float64}(256, 256, 400)
ts = Vector{Float64}(undef, 400)
\end{lstlisting}
\end{minipage}

We run this simulation until $t=500.0$ to work in a statistically equilibrated state,
and then save the result at time steps of size $0.2$.

\begin{minipage}{\textwidth}
\begin{lstlisting}
stepforward!(prob, round(Int, 500 / prob.clock.dt))
for i in 1:400
    stepforward!(prob, 20); TwoDNavierStokes.updatevars!(prob)
    vs[:,:,i] = prob.vars.v
    us[:,:,i] = prob.vars.u
    zs[:,:,i] = prob.vars.zeta
end
\end{lstlisting}
\end{minipage}

\begin{figure}
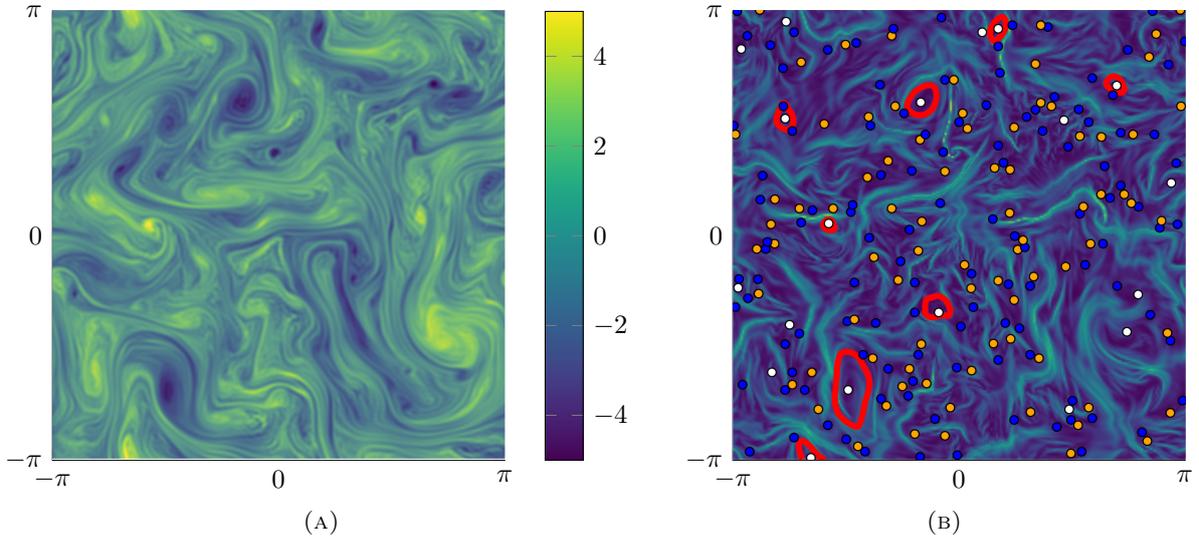

\centering
\subfloat[\label{fig:turb_vorticity}]{\input{vorticity.tex}}\quad
\subfloat[\label{fig:turb_barriers}]{\input{turb_mat_bar.tex}}
\caption{(a) Vorticity at $t=500$ in a turbulent two-dimensional velocity field.
(b) Centers of regions with index $\frac{1}{2}$ in orange, those with index $-\frac{1}{2}$ in blue, centers of elliptic regions in white.
Material barriers in red, background coloring is DBS field.}
\end{figure}

The generation of the velocity field by the above code takes just a few minutes. \Cref{fig:turb_vorticity} shows the vorticity field for a specific run at $t = 500$.

\subsubsection{Computing material barriers}

We first setup a spatially periodic interpolation of the velocity field, which is performed by the package \texttt{OceanTools.jl} \cite{OceanTools2020}.

\begin{minipage}{\textwidth}
\begin{lstlisting}
@everywhere using CoherentStructures, OceanTools
const CS = CoherentStructures
const OT = OceanTools
ts = range(0.0, step=20prob.clock.dt, length=400)
vdata = OT.ItpMetadata(xs, ys, ts, (us, vs), OT.periodic, OT.periodic, OT.flat)
\end{lstlisting}
\end{minipage}

We are now ready to compute material barriers.

\begin{minipage}{\textwidth}
\begin{lstlisting}
vortices, singularities, bg = CS.materialbarriers(
       uv_trilinear, xs, ys, range(0.0, stop=5.0, length=11),
       LCSParameters(boxradius=π/2, indexradius=0.1, pmax=1.4,
                     merge_heuristics=[combine_20_aggressive]);
       p=vdata, on_torus=true)
\end{lstlisting}
\end{minipage}

The \texttt{materialbarriers} function calculates the transport tensor field $\mathbf{T}$
used in the material-barriers approach (using finite differences for the linearized flow
map $D\flow$) and calculates material barriers. The result is shown in \cref{fig:turb_barriers}.

\begin{figure}
\centering
\input{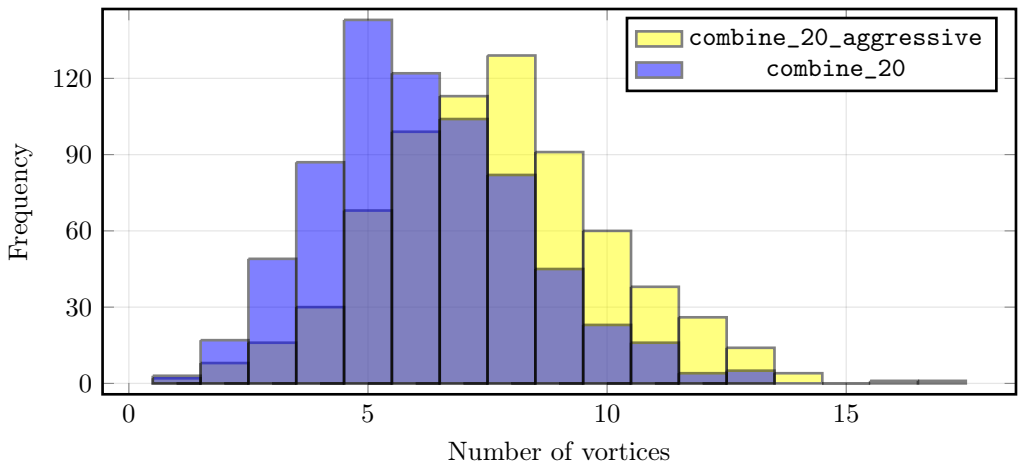}
\caption{Comparison of combination heuristics for $700$ different starting time parameters.}
\label{fig:histogram}
\end{figure}

Running with $700$ different values of $t$ took 5h 16min 26 for the less agressive heuristic,
and 8h 20min 9s for the more aggressive heuristic. \Cref{fig:histogram} shows a
histogram of the number of vortices that have been found in the 700 simulations by
the two merge heuristics; cf.~\cref{sec:mergeheuristics}. Clearly, the more
aggressive merge heuristic detects more candidate regions and, as a consequence,
more vortices. An animation containing the detected vortices over each time window
is available in the supplementary material. The animation shows some flutter in the
continuation of some vortices, especially close to their ``generation'' or
``death'', indicating room for further improvement in the robustness of the method.

\subsection{Global ocean surface flow}
\label{sec:ocean}

\begin{figure}
\centering
\subfloat[\label{fig:ocean90global}]{\includegraphics[width=.9\textwidth]{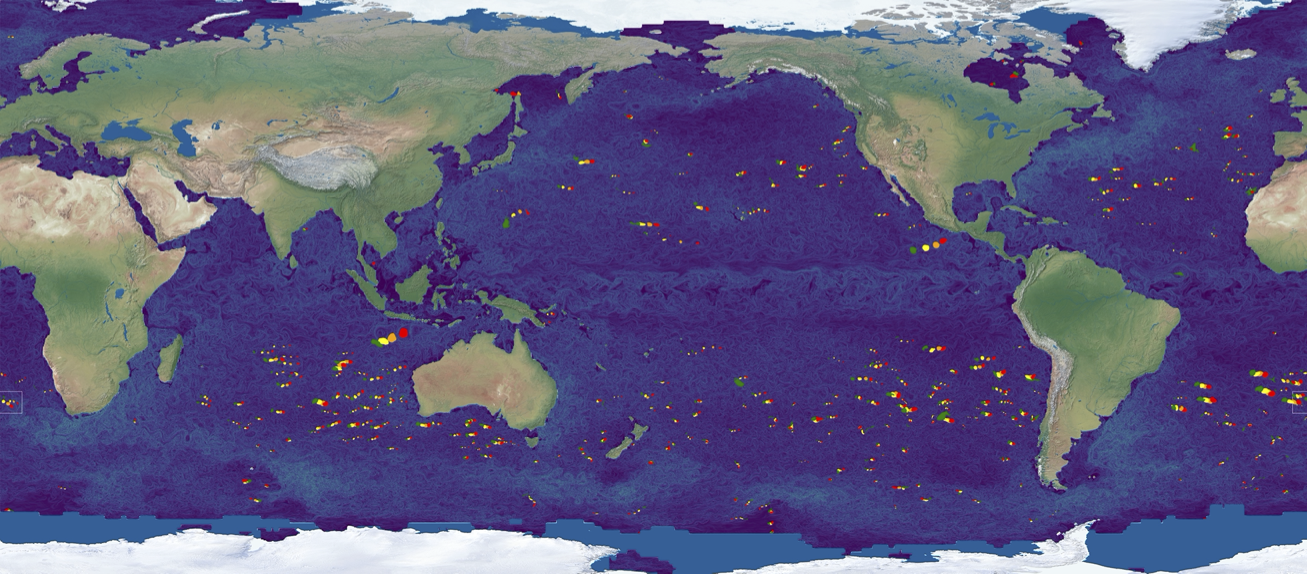}}

\subfloat[Close-up of \cref{fig:ocean90global}.]{\includegraphics[width=.9\textwidth]{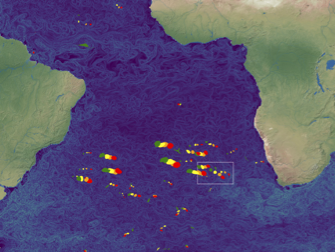}}
\caption{90-Day DBS field with filled in material diffusive transport barriers (red), their advection after 30, 60 and 90 days (orange, yellow, green respectively).}
\label{fig:ocean90}
\end{figure}

\begin{figure}
\centering
\subfloat[\label{fig:ocean30global}]{\includegraphics[width=.9\textwidth]{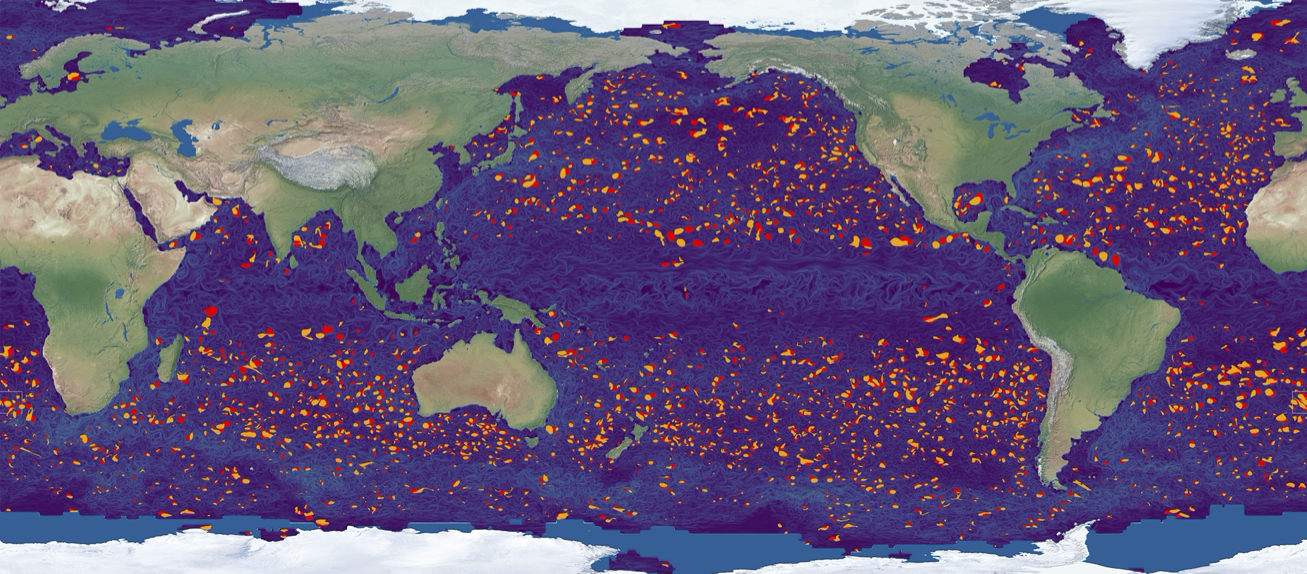}}

\subfloat[Close-up of \cref{fig:ocean30global}.]{\includegraphics[width=.9\textwidth]{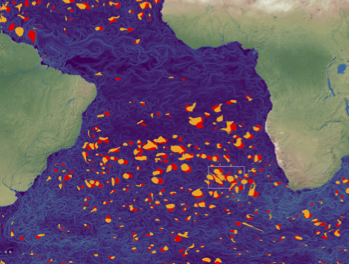}}
\caption{30-Day DBS Field with filled in material diffusive transport barriers (red), their advection after 30 days in orange.}
\label{fig:ocean30}
\end{figure}

As a very large-scale example, we compute material barriers to diffusive transport in
a global ocean surface flow. The velocity field is obtained by tricubic interpolation
(using the algorithm from \cite{Lekien2005} (cf.~ also \cite[Appendix A]{Chilenski2017}) as implemented in the \texttt{OceanTools.jl} package) of
geostrophic ocean surface velocities  from a dataset\footnote{ More specifically, from the \texttt{SEALEVEL\_GLO\_PHY\_CLIMATE\_L4\_REP\_OBSERVATIONS\_008\_057} one.} distributed by the Copernicus Marine Environment Monitoring Service.
These velocities are derived from satellite altimetry data. We choose a time window
(starting November 26, 2016) for which a small spatial piece has been used
repeatedly as a data set ``benchmark'' case in the literature; cf., for instance,
\cite{Haller2018,Froyland2018}, and earlier studies on slightly larger domains
\cite{Haller2013a,Karrasch2015}. The domain studied in the first references is
highlighted by a small white rectangle in \cref{fig:ocean30,fig:ocean90}.
\newpage
We use a regular $10000\times5000$ quadrilateral grid, i.e., with 50 million grid points.
Velocities in land-areas, or where ocean surface velocities are not available, are set
to zero. Trajectories are calculated with \texttt{CoherentStructures.jl}, which
internally uses the \texttt{DifferentialEquations.jl} package \cite{Rackauckas2017},
and relative and absolute tolerances for the ODE solver are set to $10^{-6}$. The
averaged Cauchy-Green tensor is calculated by approximating the linearization of
the flow map using finite differences, with the finite difference stencil reinitialized every 10 days. The resulting tensors are finally combined according to the product rule.

For the first test-case, we approximated the averaged, diffusion-weighted Cauchy--Green tensor for a 90-day period by averaging the Cauchy--Green tensor every 10
days from 0 to 90 days after start. The calculation of the tensors took approx.~21 hours.
We find $40,609$ $r$-stable regions, $r=0.25\degree$, of nonzero index and
obtain $6,506$ elliptic regions with the \texttt{combine\_20\_aggressive} heuristic.
The localization size $R$ is set to $2.5\degree$; cf.~\cref{sec:localization}.
This yields $357$ vortices for $\lambda$-values in $[0.7,1.4]$; the results are shown
in \cref{fig:ocean90}. The geodesic vortex computation took approx.~2.5 hours.

The second test case has an identical setup to the one described above, but with an
observation time window of only $30$ days. The computation of the tensors required approx~ 6.5 hours.
We find $44,160$ $r$-stable regions of nonzero index, with $6,399$ candidate
elliptic regions we obtained $2,255$ vortices. The results are shown in \cref{fig:ocean30}, the geodesic vortex computation took again roughly 2.5 hours.

While we are not qualified to interpret \cref{fig:ocean30,fig:ocean90} from an
oceanographic point of view, we do note that the region of known active vortex
generation to the west of the southern tip of Africa due to Agulhas leakage \cite{Ruijter1999} can be nicely recognized.

\begin{figure}
\centering
\includegraphics[width=.9\textwidth]{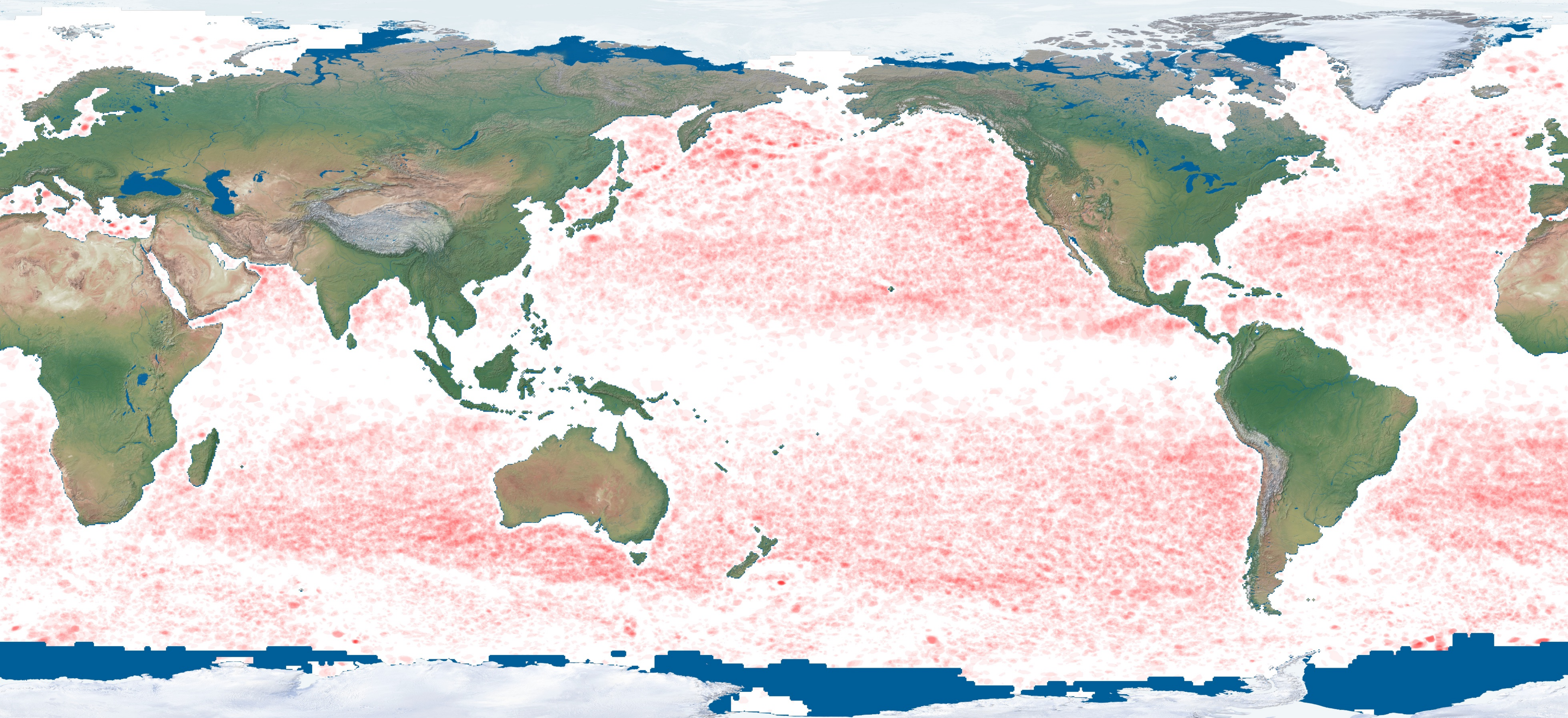}
\caption{Coherent Lagrangian vortices from 36 30-day simulations, covering
roughly 3 years. Every coherent vortex is shown in transparent red, with several
spatially overlapping vortices increasing the color intensity.}
\label{fig:vortex_density}
\end{figure}

To further demonstrate the capabilities of our method, we have run 36 30-day
simulations with a 30-day time lag, i.e., spanning a period of approximately 3 years.
We then plot the initial coherent vortex locations superimposed onto each other with
a (uniform) transparency in \cref{fig:vortex_density}. This visualization highlights regions where
vortices tend to occur more often. \Cref{fig:vortex_density} is part of ongoing work on the statistics of
coherent transport in the global ocean.

\section{Conclusion}

In this work, we have described in detail our implementation of the index-theory-based approach \cite{Karrasch2015}
to computing coherent Lagrangian vortices including material barriers to diffusive transport
as closed null-geodesics of some Lorentzian metric tensor field. Despite the fact that
there is no theoretical guarantee that all closed null-geodesics are found with the currently implemented merge heuristics,
the approach is generally well-suited for genuinely challenging flow problems. This is due to the fact
that while the effort for the computations related to the index-theory-based
determination of candidate regions is basically negligible, its outcome
dramatically reduces computational effort in the closed orbit detection via localization.
In fact, there are cases (especially when reducing the number of grid points) where we do find \emph{more}
vortices with our implementation than with the implementation described in
\cite{Serra2017}. We suspect that this is because of the high degree of accuracy and resolution required in \cite{Serra2017}.
We have demonstrated our method on two test cases of considerable size, yet with
affordable computational resources.
Comparable simulation studies have, to the best of our knowledge and
with the possible exception of \cite{Abernathey2018}, not been published before.

In this paper, we have restricted ourselves to presenting the theory and the
implementation for computing closed null-geodesics of tensor fields. The exact
same considerations with minor modifications carry over to the case of
\emph{constrained diffusion barriers}, as introduced in \cite{Haller2019} as closed
orbits of the \emph{transport vector field}. The corresponding implementation
building on the same machinery described here is available in \texttt{CoherentStructures.jl}.

When applied to such challenging problems, remaining issues become apparent,
indicating room for further improvement. Interesting new features that are work in
progress or under consideration include the approximation of the tensor field
$\mathbf{T}$ from scattered trajectories in order to have a purely data-driven
approach for calculating coherent Lagrangian vortices, the use of boundary value
problem solvers or topological methods à la \cite{Wischgoll2001} for the actual
closed orbit detection, and other miscellaneous optimizations. In any case, our
software, even in its current form, paves the way to global transport studies on the ocean surface from a
Lagrangian viewpoint, which we hope may be of interest to the oceanographical community.

\appendix

\section{Index theory for planar line fields}
\label{app:index_theory}

We give a brief summary of some basic results from the classic index theory of planar line fields.
For more details (and proofs connected to the classical theory), we refer the reader
to \cite{Perko2001} and \cite{Hopf1989} for the vector field and line field cases,
respectively. We outline the theory for fields defined on an open, simply connected subset $\Omega$ of
$\R^2$, but it can be easily extended to more general Riemannian
manifolds by viewing $\Omega$ as the domain of a manifold parametrization.
Notably, indices are independent of the specific Riemannian metric structure \cite[Ch.~3, Thm.~1.4]{Hopf1989}.

First, recall that a \emph{Jordan curve} is a continuous function $\gamma\colon [a,b]\subset\R \to \Omega$
that is closed ($\gamma(a) = \gamma(b)$) and simple, i.e., it does not self-intersect.
Such curves decompose $\Omega$ into three sets: the curve itself (the image of
$\gamma$), its interior, and its exterior (which is unbounded if $\Omega=\mathbb{R}^2$
or touches the boundary of $\Omega$ otherwise).

A line field on a domain $D \subset \Omega \subset \R^2$ is
a continuous map $\lf\colon D \rightarrow \mathbb P^1$, where $\mathbb P^1$ is the one-dimensional real projective space, i.e the set of all one-dimensional subspaces of $\R^2$.
The real projective space can be given the structure of a compact, smooth manifold,
and can be parametrized by an angle coordinate as follows: assign to any one-dimensional vector space $V$ the angle between (any vector of) the subspace and the $x$-axis.
This assignment is unique up to a multiple of $\pi$, taking into account the two possible orientations within $V$. We assume that $\lf$ is only defined on $D$ and not on the whole of $\Omega$ due to the existence of singularities, see below.
\newpage
\begin{definition}[Line-field index of a curve]\label{def:lf_index}
Let $\gamma$ be a Jordan curve in $D$, and $\theta$ be a continuous function giving the angle between $\lf(\gamma(s))$ and the $x$-axis, i.e., this means $\lf(\gamma(s)) = \spn\lbrace (\cos\theta, \sin\theta)^T\rbrace $ for $s\in [a,b]$. Then the \emph{index of $\gamma$ relative to $\lf$} is given by
\begin{align}\label{eq:lfindex}
\ind_\lf(\gamma) \coloneqq \frac{1}{2\pi} (\theta(b) - \theta(a)) = \frac{1}{2\pi}\int_{\gamma} \,\mathrm{d}\theta,
\end{align}
where, for a piecewise $C^1$-curve $\gamma$, $\mathrm{d}\theta$ is the derivative of the enclosed angle $\theta$ along $\gamma$.
\end{definition}

\begin{figure}
\centering
\subfloat[Two curves $\gamma_1$ (left, black) and $\gamma_2$ (right, blue).]{
\begin{tikzpicture}[]
\begin{axis}[height = 0.4\linewidth, ylabel = {}, axis lines=none,
xmin = {-0.09115261837150107}, xmax = {2.091152618371501}, ymin = {-0.06115261837150108}, ymax = {1.0611526183715012}, xlabel = {}, unbounded coords=jump,scaled x ticks = false,xlabel style = {font = {\fontsize{11 pt}{14.3 pt}\selectfont}, color = black, draw opacity = 1.0, rotate = 0.0},xmajorgrids = true,xtick = {0.0,0.5,1.0,1.5,2.0},xticklabels = {$0.0$,$0.5$,$1.0$,$1.5$,$2.0$},xtick align = inside,xticklabel style = {font = {\fontsize{8 pt}{10.4 pt}\selectfont}, color = black, draw opacity = 1.0, rotate = 0.0},x grid style = {color = black,
draw opacity = 0.1,
line width = 0.5,
solid},axis lines* = left,x axis line style = {color = black,
draw opacity = 1.0,
line width = 1,
solid},scaled y ticks = false,ylabel style = {font = {\fontsize{11 pt}{14.3 pt}\selectfont}, color = black, draw opacity = 1.0, rotate = 0.0},ymajorgrids = true,ytick = {0.0,0.25,0.5,0.75,1.0},yticklabels = {$0.00$,$0.25$,$0.50$,$0.75$,$1.00$},ytick align = inside,yticklabel style = {font = {\fontsize{8 pt}{10.4 pt}\selectfont}, color = black, draw opacity = 1.0, rotate = 0.0},y grid style = {color = black,
draw opacity = 0.1,
line width = 0.5,
solid},axis lines* = left,y axis line style = {color = black,
draw opacity = 1.0,
line width = 1,
solid},    xshift = 0.0mm,
    yshift = 0.0mm,
    ]

\addplot+ [color = black,
line width = 2,
solid,mark = none,
forget plot]coordinates {
(0.0, 0.0)
(0.98, 0.0)
(0.98, 1.0)
(0.0, 1.0)
(0.0, 0.0)
};
\addplot+ [color = blue,
line width = 2,
solid,mark = none,
forget plot]coordinates {
(1.02, 0.0)
(2.0, 0.0)
(2.0, 1.0)
(1.02, 1.0)
(1.02, 0.0)
};
\addplot+ [color = black,
draw opacity = 1.0,
line width = 2,
solid,mark = none, ->,
forget plot]coordinates {
(0.0, 1.0)
(0.0, 0.5)
};
\addplot+ [color = black,
draw opacity = 1.0,
line width = 2,
solid,mark = none, ->,
forget plot]coordinates {
(0.0, 0.0)
(0.5, 0.0)
};
\addplot+ [color = black,
draw opacity = 1.0,
line width = 2,
solid,mark = none, ->,
forget plot]coordinates {
(0.98, 0.0)
(0.98, 0.5)
};
\addplot+ [color = black,
draw opacity = 1.0,
line width = 2,
solid,mark = none, ->,
forget plot]coordinates {
(0.98, 1.0)
(0.5, 1.0)
};
\addplot+ [color = blue,
draw opacity = 1.0,
line width = 2,
solid,mark = none, ->,
forget plot]coordinates {
(1.02, 0.0)
(1.5, 0.0)
};
\addplot+ [color = blue,
draw opacity = 1.0,
line width = 2,
solid,mark = none, ->,
forget plot]coordinates {
(2.0, 0.0)
(2.0, 0.5)
};
\addplot+ [color = blue,
draw opacity = 1.0,
line width = 2,
solid,mark = none, ->,
forget plot]coordinates {
(2.0, 1.0)
(1.5, 1.0)
};
\addplot+ [color = blue,
draw opacity = 1.0,
line width = 2,
solid,mark = none, ->,
forget plot]coordinates {
(1.02, 1.0)
(1.02, 0.5)
};
\end{axis}

\end{tikzpicture}
}
\subfloat[The sum $\gamma_1 + \gamma_2$.]{
\begin{tikzpicture}[]
\begin{axis}[height = 0.4\linewidth, ylabel = {}, axis lines=none,
xmin = {-0.09115261837150107}, xmax = {2.091152618371501}, ymin = {-0.06115261837150108}, ymax = {1.0611526183715012},
xlabel = {}, unbounded coords=jump,scaled x ticks = false,xlabel style = {font = {\fontsize{11 pt}{14.3 pt}\selectfont}, color = black, draw opacity = 1.0, rotate = 0.0},xmajorgrids = true,xtick = {0.0,0.5,1.0,1.5,2.0},xticklabels = {$0.0$,$0.5$,$1.0$,$1.5$,$2.0$},xtick align = inside,xticklabel style = {font = {\fontsize{8 pt}{10.4 pt}\selectfont}, color = black, draw opacity = 1.0, rotate = 0.0},x grid style = {color = black,
draw opacity = 0.1,
line width = 0.5,
solid},axis lines* = left,x axis line style = {color = black,
draw opacity = 1.0,
line width = 2,
solid},scaled y ticks = false,ylabel style = {font = {\fontsize{11 pt}{14.3 pt}\selectfont}, color = black, draw opacity = 1.0, rotate = 0.0},ymajorgrids = true,ytick = {0.0,0.25,0.5,0.75,1.0},yticklabels = {$0.00$,$0.25$,$0.50$,$0.75$,$1.00$},ytick align = inside,yticklabel style = {font = {\fontsize{8 pt}{10.4 pt}\selectfont}, color = black, draw opacity = 1.0, rotate = 0.0},y grid style = {color = black,
draw opacity = 0.1,
line width = 0.5,
solid},axis lines* = left,y axis line style = {color = black,
draw opacity = 1.0,
line width = 2,
solid},    xshift = 0.0mm, yshift = 0.0mm,
legend style = {color = black,
draw opacity = 1.0,
line width = 2,
solid,fill = white,font = {\fontsize{8 pt}{10.4 pt}\selectfont}},colorbar style={title=}]

\addplot+ [color = black,
draw opacity = 1.0,
line width = 2,
solid,mark = none,
mark size = 2.0,
mark options = {
    color = black, draw opacity = 1.0,
    fill = black, fill opacity = 1.0,
    line width = 2,
    rotate = 0,
    solid
},forget plot]coordinates {
(0.0, 0.0)
(2.0, 0.0)
(2.0, 1.0)
(0.0, 1.0)
(0.0, 0.0)
};
\addplot+ [color = black,
draw opacity = 1.0,
line width = 2,
solid,mark = none, ->,
forget plot]coordinates {
(0.0, 1.0)
(0.0, 0.5)
};
\addplot+ [color = black,
draw opacity = 1.0,
line width = 2,
solid,mark = none, ->,
forget plot]coordinates {
(2.0, 1.0)
(1.0, 1.0)
};
\addplot+ [color = black,
draw opacity = 1.0,
line width = 2,
solid,mark = none, ->,
forget plot]coordinates {
(2.0, 0.0)
(2.0, 0.5)
};
\addplot+ [color = black,
draw opacity = 1.0,
line width = 2,
solid,mark = none, ->,
forget plot]coordinates {
(0.0, 0.0)
(1.0, 0.0)
};
\end{axis}

\end{tikzpicture}
}
\caption{Addition of curves: segments that are traversed in opposing directions annihilate each other in the sum.}
\label{fig:addcurves}
\end{figure}
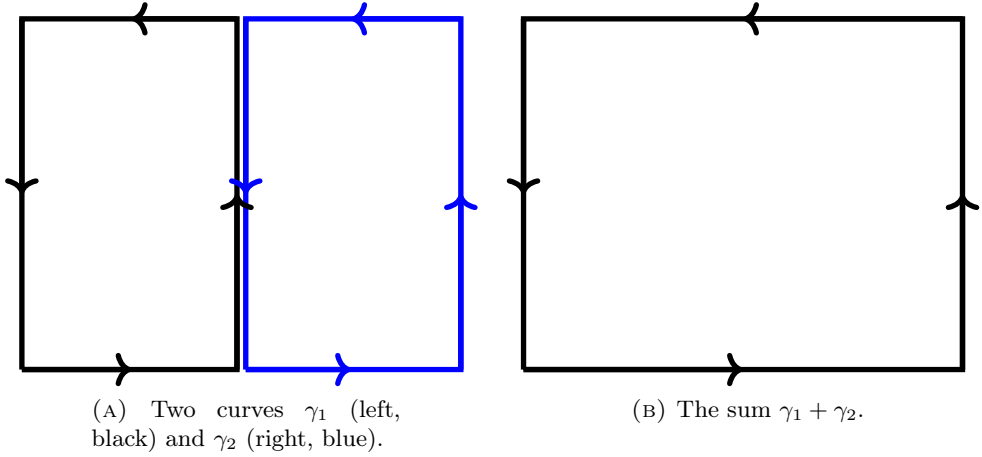

Clearly, for $\ind_{\lf}(\gamma)$
to be defined, $\gamma$ must not pass through points in $\Omega \setminus D$. The index of $\gamma$
is invariant under orientation-preserving reparametrizations, and has the opposite signs under
orientation-reversing reparametrizations. We can therefore identify $\gamma$ with its image
$\gamma([a,b])$, fixing its orientation. Moreover, the index is (i) invariant under homotopies, i.e.,
continuous deformations of $\gamma$ in $D$, and (ii) additive under
composition of curves in the sense that if $\gamma_1$, $\gamma_2$ are two closed curves
and $\gamma = \gamma_1 + \gamma_2 \coloneqq \overline{(\gamma_1 \cup \gamma_2) \setminus (\gamma_1 \cap \gamma_2)}$
is again a closed curve (see \cref{fig:addcurves}), then
\[
\ind_{\lf}(\gamma) = \ind_{\lf}(\gamma_1) + \ind_{\lf}(\gamma_2),
\]
provided that each index is defined. We may also define an index for subsets of $\Omega$
whose boundaries are Jordan curves (in $D$) or finite disjoint unions of such subsets, the additivity of the index then extends to additivity
under disjoint unions, and unions that are disjoint except for a common segment of a Jordan curve on each boundary.

\begin{definition}[Index of a region]
Let $R \subset \Omega$ be the interior of a (positively oriented)
Jordan curve $\gamma$ with values in $D$.
Then we define the \emph{index of $R$ (relative to $\lf$)} as
\[
\ind_{\lf}(R) \coloneqq \ind_{\lf}(\gamma).
\]
\end{definition}

\begin{definition}
A \emph{closed orbit} of a line field $\lf\colon D \rightarrow \mathbb P^1$ is defined as a smooth curve $\gamma\colon [a,b] \rightarrow D$ which is
\begin{enumerate}
\item closed: $\gamma(a) = \gamma(b)$, and
\item an integral curve, i.e., $\gamma'(s) \in \lf(\gamma(s))$ for $s \in [a,b]$.
\end{enumerate}
\end{definition}

Since tangent vectors along closed orbits turn once, we have the following well-known property.

\begin{theorem}
The index of a closed orbit of $\lf$ relative to $\lf$ is $1$.
\end{theorem}

There may exist points in $\Omega$ at which a line field $\lf$ is not defined and,
importantly, to which it cannot be extended continuously. Such points are referred to
as \emph{(line field) singularities}. As with critical points of vector fields, the angle
$\theta$ in \cref{eq:lfindex} is not well-defined at singularities and hence Jordan curves must not pass through singularities for their index to be well-defined.
If a curve $\gamma$ encloses no singularities, then it must have index $0$. The theorem above hence shows that closed orbits of line-fields must contain at least one singularity.

\begin{definition}[Line-field index of a region]
Let $\gamma$ be an anticlockwise Jordan curve in $D$ whose interior (with respect to $\Omega$) is a set $R$. Then we define
\[
\ind_\lf(R) \coloneqq \ind_\lf(\gamma).
\]
\end{definition}

\begin{definition}
We call a set $R$ with index 1 (with respect to $\lf$) \emph{elliptic (with respect to $\lf$)}.
\end{definition}

We note in passing that if a singularity $p$ of $\lf$ is isolated, we can assign an index to it
as the index of an arbitrary small region containing $p$. Therefore, if a region $R$ contains
finitely many isolated singularities, index additivity implies that the index of $R$
is the sum of the indices of its enclosed singularities.

Clearly, every vector field $v$ gives rise to a line field by applying the $\spn$
operation at each point $x\in\Omega$, but the converse is not true in general: the
obstruction comes exactly from line field singularities.
For a line field $\lf=\spn{v}$ induced by a vector field $v$, singularities correspond
exactly to critical points of $v$, and, moreover, their indices w.r.t.~$v$ and $\lf$ coincide.
Note, however, that in general the line field index is an element of the set of half-integers
$\frac12\mathbb{Z}$, in contrast to vector field indices that take only integer values.
This shows that half-integer singularities form an obstruction to representing a line field $\lf$
as a vector field locally.

\section{Applications of index theory to calculating geodesic vortices}
\label{app:applications}

Geodesic vortices are closed orbits of the line-field $\eta^\pm_\lambda$. By the theory outlined above,
it follows that if $\gamma$ is a geodesic vortex, it must have index $1$ relative to $\eta^\pm_\lambda$.
Geodesic vortices thus enclose elliptic regions (with respect to $\eta^\pm_\lambda$).

Following the approach in \cite{Karrasch2015}, we can continuously deform the line-field
$\eta^\pm_\lambda$ to the line-field $\xi_1$, and continuity of the index (and the fact that
it only takes half-integer values) implies that the index of $\gamma$ relative to $\xi_1$
must also be $1$. Therefore geodesic vortices enclose elliptic regions with respect to $\xi_1$.
Changing the line field of reference has the following advantage: depending on the parameter value $\lambda$,
$\eta^\pm_\lambda$ may not be defined in open regions, as opposed to $\xi_1$, which---for
generic tensor fields---is defined everywhere except at isolated singularities; cf.~\cite[Sect.~4(b)]{Karrasch2015}.

\section*{Acknowledgements}
This study has been conducted using E.U. Copernicus Marine Service Information.
We wish to thank Navid Constantinou for technical help regarding the usage of the
\texttt{GeophysicalFlows.jl} package, and Tom Patterson, \url{www.shadedrelief.com},
for providing the global Earth map that we have used in \cref{sec:ocean}.

\bibliographystyle{plainurl}

\begin{small}

\end{small}

\end{document}